\documentclass[10pt]{article}

\usepackage{amssymb}
\usepackage{amsmath}
\usepackage{amsthm} 
\usepackage{mathrsfs}                                                           
\usepackage{slashbox}  
\usepackage[all]{xy}
\usepackage{graphicx}
\usepackage{wrapfig}

\def\2{{1\over 2}}

\newcommand{\rf}[1]{(\ref{#1})}
\def\b{\bar}

\newcommand{\ud}{\mathrm{d}}
\renewcommand{\t}{\tilde}
\newcommand{\p}{\partial}
\newcommand{\bp}{\bar{\partial}}

\newcommand{\mA}{\mathbf{A} }
\newcommand{\mB}{\mathbf{B} }
\newcommand{\mC}{\mathbf{C} }
\newcommand{\mD}{\mathbf{D} }
\newcommand{\mV}{\mathbf{V} }
\newcommand{\mW}{\mathbf{W} }

\newcommand{\mF}{\mathcal{F} }
\newcommand{\mQ}{\mathcal{Q} }
\newcommand{\m}{\ud*\ud}

\newcommand{\mg}{\mathfrak{g}}
\newcommand{\ma}{\mathfrak{a}}
\newcommand{\mb}{\mathfrak{b}}
\newcommand{\mv}{\mathfrak{v}}
\newcommand{\mau}{\mathfrak{u}}
\newcommand{\mG}{\mathfrak{G}}
\newcommand{\mK}{\mathfrak{K}}
\title{
\bf{Conformal Field Theory and Algebraic Structure of Gauge Theory}}
\author{Anton M. Zeitlin\footnote{anton.zeitlin@yale.edu http://math.yale.edu/$\sim$az84, http://www.ipme.ru/zam.html} \\
Department of Mathematics,\\
Yale University,\\
442 Dunham Lab, 10 Hillhouse Ave\\
New Haven, CT 06511}
\date{}

\begin{document}
\maketitle
\begin{abstract}
We consider various homotopy algebras related to Yang-Mills theory and two-dimensional conformal field theory (CFT). Our main objects of study are Yang-Mills $L_{\infty}$ and $C_{\infty}$ algebras and their relation to the certain algebraic structures of Lian-Zuckerman type in CFT. We also consider several examples of algebras related to gauge theory, involving first order formulations and gauge theories with matter fields.
\end{abstract}
\section{Introduction}
The homotopy algebras ($A_{\infty}$, $C_{\infty}$, $L_{\infty}$, etc) drew a lot of attention in many areas of String Theory and Field Theory. 
In Field Theory, they are known to appear, e.g. when one considers Batalin-Vilkovisky (BV) quantization \cite{bv}-\cite{schkonts} of the theory.  Similar constructions appear in case of String Field Theory (see e.g. \cite{zwiebach}, \cite{stakaj}) and in Topological String Theory (see e.g. \cite{herbst}) as a necessary ingredient. 
The main goal of the paper is to find interrelations between a natural
homotopy algebraic structure of conformal field theory, constructed by
B. Lian and G. Zuckerman, and certain algebraic structures in gauge
theory.

In \cite{lz} it was shown that the chiral BRST complex possesses a bilinear  operation which leads to the 
homotopy commutative associative algebra. Moreover, it appears that using this operation one can build an odd (w.r.t. the ghost number) bilinear operation which satisfies the relations of a homotopy Gerstenhaber algebra ($G_{\infty}$). In this paper, we consider the CFT of open string on a half-plane and show that adapting the Lian-Zuckerman construction to this case, one can reproduce  the $C_{\infty}$ algebra of Yang-Mills (YM) theory, which we also present in this paper. We also make a conjecture about the possible underlying $G_{\infty}$ algebra structure in YM theory. 

In \cite{ym}, \cite{bvym} we showed that Yang-Mills equations considered on a flat space, 
coincide on the formal level with the generalized Maurer-Cartan (GMC) equation for a certain $L_{\infty}$ algebra, 
such that the corresponding (formal) gauge symmetries coincide with natural symmetries of the GMC equation. 
In section 2 of this paper, we generalize these constructions to the case of general Riemannian manifold. We demonstrate that one can interpret YM connections, i.e. connections satisfying YM equation, as the solution of the Maurer-Cartan equation for a sheaf of $L_{\infty}$ algebras. We show the relation of this algebraic structures to the detour complex \cite{detour} and define certain graded antisymmetric multilinear products, which  lead to the BV Yang-Mills action written in a homotopy Chern-Simons form. 
 
Section 3 is devoted to the main topic of the paper. In subsection 3.1., we describe the $C_{\infty}$ algebra which is underlying for the $L_{\infty}$ algebra of YM theory. We also show, how one can embed the complex on which the YM $C_{\infty}$ algebra is based, into the BRST complex of the open string theory. 
Then, via point-splitting regularization, the Lian-Zuckerman bilinear operation is generalized in order to be applied to open string states. However, the generalization should be modified further if one wants to get
a homotopy algebra on all the BRST complex, since in these
considerations logarithms in operator product expansions,  which
correspond to higher order $\alpha'$ corrections, are not taken into
account (since we are interested in the lowest ones).
At the same time, rewriting via point-splitting the usual "chiral"
Lian-Zuckerman product one can get an interesting generalization of this
bilinear operation which is defined in subsection 3.4 and will
be studied further  in \cite{zeitnew}. 
We also claim that the algebraic structures we introduced should produce the underlying  symmetry of the perturbation series (see also \cite{zeit3}) corresponding to the open  string sigma models, renormalized appropriately (via carefully defined point-splitting regularization). 
In this article we study only the lowest order $\alpha'$-corrections to the Lian-Zuckerman products. 
 The resulting algebra on a certain subcomplex of the
BRST complex is shown to be quasiisomorphic to the algebra which contains
the YM $C_{\infty}$algebra as a subalgebra. It is worth noting here, that in \cite{fz} the BRST complex and the Lian-Zuckerman construction were used to reproduce another noncommutative algebraic structure, namely, the Quantum Group $SL_q(2)$.

The last part of this section is two-folded. Firstly, we give the "cohomological" version of gauge fixing in YM (subsection 3.6), which extends the amount of examples of theories which allow such type of gauge fixing (and allows to apply the formalism of \cite{costello} to the ordinary YM theory, avoiding first order formulations). Secondly, the operator used in gauge fixing is induced from the one, which 
creates a (coboundary) homotopy Gerstenhaber algebra in Lian-Zuckerman constructions. Therefore 
we make a conjecture (see subsection 3.7), that this operator generates a homotopy Gerstenhaber algebra associated with YM theory. 
 
Section 4 is devoted to various examples of algebras related to YM theory. The first example comes from the first order formulation of YM: its underlying algebraic structure is not homotopic, but just graded Abelian associative algebra with a differential. Next, we switch to  dimension 4 and consider the Abelian algebra related to another first order formulation (it was also studied in \cite{costello}). We find that there exists another algebra containing both this 
Abelian algebra and the YM $C_{\infty}$ algebra as subalgebras. This extended algebra appears to be related to Topological YM theory \cite{blau}. In subsection 4.3, we show how to add matter fields to the theory and interpret Higgs mechanism as a deformation of a differential in the given $L_{\infty}$ algebra.  Further directions of study are briefly sketched  in Final Remarks.

\section{Yang-Mills $L_{\infty}$ algebra}
{\bf 2.0. Notation and conventions.}
 Throughout the paper, $M$ denotes a smooth Riemannian manifold and $D$ denotes its dimension. 
 
 Let $\mA$ be a connection 1-form for some principal bundle over a Riemannian manifold M. Then the Yang-Mills action functional is:
\begin{eqnarray}
S_{YM}=\frac{1}{2}Tr\int_M(\mathbf{F}\wedge *\mathbf{F}), \quad \mathbf{F}=\ud \mA+\mA\wedge \mA,
\end{eqnarray}
by $Tr$ we denote the canonical invariant form for the Lie algebra, where $\mA$ takes values in.
 
The YM equations of motion are:
\begin{eqnarray}\label{yme}
\frac{\delta S_{YM}}{\delta\mA}=\ud_{\mA}*\mathbf{F}=0,
\end{eqnarray}
where $\ud_{\mA}$ is an exterior covariant derivative associated with the connection $\mA$.\\
The equations \rf{yme} possess a local infinitesimal symmetry
\begin{eqnarray}
\mA\to \mA+\epsilon(\ud u+[\mA,u]),
\end{eqnarray}
where u is a function with values in a Lie algebra and $\epsilon$ is an infinitesimal parameter. 

Consider the smooth Riemannian manifold $M$ and the 
following chain complex:
\begin{eqnarray}\label{maxwell}
0\xrightarrow{ }\Omega^{0}(M)\xrightarrow{\ud}\Omega^{1}(M)
\xrightarrow{\m}\Omega^{D-1}(M)\xrightarrow{\ud}\Omega^{D}(M)\to 0,
\end{eqnarray}
where $\Omega^{k}(M)$ denotes the space of $k$-forms on the manifold $M$. 
In this paper, we will call \rf{maxwell} the {\it Maxwell complex} and denote as $(\mF^{\bf \cdot}, \mQ)$:
\begin{eqnarray}
0\to\mathcal{F}^{0}\xrightarrow{\mathcal{Q}}\mathcal{F}^{1}\xrightarrow{
\mathcal{Q}}
\mathcal{F}^{2}\xrightarrow{\mathcal{Q}}\mathcal{F}^{3}\to 0.
\end{eqnarray}
Here the differential is denoted by $\mQ$, and $\mathcal{F}^{i}$ stands for the appropriate space of differential forms 
according to grading. 
In the following, we will refer to the grading in this complex by means of letter $n$, namely, for element $f\in \mathcal{F}^{k}$, $n_f\equiv k$. 
For our purposes it will be useful to tensor it with some reductive Lie algebra $\mg$. We will denote the resulting complex $(\mF_{\mg}^{\bf \cdot}, \mQ)$.  
 One can also make an extension of the complex above by means of the contractible complex 
\begin{eqnarray}
0\to \mathfrak{G}^1\xrightarrow{\mQ}\mathfrak{G}^2\to 0
\end{eqnarray}
such that  $\mathfrak{G}^1= \Omega^0(M)$,  $\mathfrak{G}^2= \Omega^D(M)$ and $\mQ$ acts as the Hodge star. Namely, we consider
\begin{eqnarray}
\mF^{0}_{ext}=\mF^{0},\quad \mF^{1}_{ext}=\mF^1\oplus\mathfrak{G}^{1}, \quad \mF^{2}_{ext}=\mF^{2}\oplus\mathfrak{G}^{2},\quad \mF^{3}_{ext}=\mF^{3}.
\end{eqnarray}
Throughout the paper we will refer to the resulting complex $(\mF_{ext}^{\bf \cdot}, \mQ)$ as the $extended$ Maxwell complex.\\

\noindent{\bf 2.1. YM $L_{\infty}$ algebra.} We define bilinear and trilinear operations
\begin{eqnarray}
&&[\cdot, \cdot]_h: \mathcal{F}^i_{\mathfrak{g}}\otimes \mathcal{F}^j_{\mathfrak{g}}\to \mathcal{F}^{i+j}_{\mathfrak{g}},\\
&&\label{3lin}[\cdot, \cdot, \cdot]_h: \mathcal{F}^i_{\mathfrak{g}}\otimes \mathcal{F}^j_{\mathfrak{g}}\otimes \mathcal{F}^k_{\mathfrak{g}}\to 
\mathcal{F}^{i+j+k-1}_{\mathfrak{g}},
\end{eqnarray}
which are respectively graded antisymmetric bilinear and 3-linear operations (here obviously, 
$\mathcal{F}^i_{\mathfrak{g}}=0$ for $i<0$ and $i>3$). 
The values of the bilinear operation $[f_1,f_2]_h$ for $f_{1,2}\in \mathcal{F}^{\mathbf{\cdot}}_{\mathfrak{g}}$ are defined by means of the following table: \\

$[f_1,f_2]_h$=
\begin{tabular}{|l|c|c|c|r|}
\hline
\backslashbox{$ f_2$}{$f_1$}&\makebox{$v$} & \makebox{$\mA$} & \makebox{$\mV$} & \makebox{$a$} \\
\hline
$w$ & $[v,w]$& $[\mA,w]$& $[\mV,w]$&$[a,w]$\\
\hline
$\mB$ &$[v,\mB]$  & $\{\mA,\mB\}$   &   $-[\mB,\mV]$&0\\
\hline
$\mW$ & $[v,\mW]$ &  $ [\mA,\mW]$ &    0&0\\
\hline
$b$ & $[v,b]$ &   0 &    0&0\\
\hline
\end{tabular}\\

\noindent Here	 $f_1$ takes values in the set $\{v, \mA, \mV, a\}$ from the first row, $f_2$ takes values in the set $\{w, \mB, \mW, b\}$. Other elements in the table represent the value of bilinear operation $[f_1,f_2]_h$ for appropriately chosen $f_1$ and $f_2$. In the table above $v,w\in\mathcal{F}^0_{\mathfrak{g}}$; $\mA,\mB\in \mathcal{F}^1_{\mathfrak{g}}$; 
$\mV,\mW\in\mathcal{F}^2_{\mathfrak{g}}$; $a,b\in\mathcal{F}^3_{\mathfrak{g}}$. 
The bilinear operation $\{\mA,\mB\}$ is defined as follows:
\begin{eqnarray}
&&\{\mA,\mB\}=[\mB,*\ud\mA]+[\mA,*\ud\mB]+\ud *[\mA, \mB].
\end{eqnarray}
The operation \rf{3lin} is defined to be nonzero only when all the arguments belong to $\mF^1$. For $\mA$, 
$\mB$, $\mC$$\in \mF^1$ we have: 
\begin{eqnarray}
[\mA,\mB ,\mC ]_h=[\mA,*[\mB,\mC]]+[\mC,*[\mA,\mB]]+[\mB,*[\mC,\mA]].
\end{eqnarray}

\vspace{3mm}

We claim that the graded 
antisymmetric multilinear 
operations, defined above, satisfy the relations of a homotopic Lie algebra. Namely, the following proposition holds.

\vspace{3mm}

\noindent{\bf Proposition 2.1.} {\it Let $a_1,a_2, a_3, b, c$ $\in$ $\mF$. Then the following relations hold:
\begin{eqnarray}
&&\mQ[a_1,a_2]_h=[\mQ a_1,a_2]_h+(-1)^{n_{a_1}}[a_1,\mQ a_2]_h,\nonumber\\
&&\mQ[a_1,a_2, a_3]_h+[\mQ a_1,a_2, a_3]_h+(-1)^{n_{a_1}}[a_1,\mQ a_2, a_3]_h+\nonumber\\
&&(-1)^{n_{a_1}+n_{a_2}}[ a_1, a_2, \mQ a_3]_h+[a_1,[a_2, a_3]_h]_h-[[a_1,a_2]_h, a_3]_h-\nonumber\\
&&(-1)^{n_{a_1}n_{a_2}}[a_2,[a_1, a_3]_h]_h=0,\nonumber\\
&&[b,[a_1,a_2, a_3]_h]_h-(-1)^{n_b(n_{a_1}+n_{a_2}+n_{a_3})}[a_1,[a_2, a_3, b]_h]_h+\nonumber\\
&&(-1)^{n_{a_2}(n_{b}+n_{a_1})}[a_2,[b,a_1, a_3]_h]_h-\nonumber\\
&&(-1)^{n_{a_3}(n_{a_1}+n_{a_2}+n_{b})}
[a_3,[b, a_1,a_2]_h]_h\nonumber\\
&&=[[b,a_1]_h,a_2, a_3]_h+(-1)^{n_{a_1}n_{b}}[a_1,[b,a_2]_h, a_3]_h+\nonumber\\
&&(-1)^{(n_{a_1}+n_{a_2})n_{b}}[a_1,a_2, [b,a_3]_h]_h,\nonumber\\
&&[[a_1,a_2, a_3]_h,b,c]_h=0.
\end{eqnarray} }
The proof can be found in \cite{ym}.\\
%%%%%%%%%%%%%%%%%%%%%%%%%%%%%%%%%%%%%%%%%%%%%%%%%%%%%%%%%%%%%%%%%%%%%%%%%%%
%{\bf Remark.} 
%Let's denote $d_0=\mQ$, $d_1=[\cdot, \cdot]_h$, $d_2=[\cdot, \cdot, \cdot]_h$. Then the relations \rf{rel} together with condition $\mQ^2=0$ can be summarized 
%in the following way: 
%\begin{eqnarray}\label{d2}
%D^2=0,
%\end{eqnarray}
%where $D=d_0+\theta d_1+\theta^2d_2$ . Here, $\theta$ is some formal parameter anticommuting with $d_0$ and $d_2$. 
%We remind that $d_0$ raises grading number by $1$, $d_1$ leaves it unchanged while $d_2$ lowers grading by 1. Therefore, $d_0,d_2$ are odd elements as well as the parameter $\theta$, but $d_1$ is even. Therefore, \rf{d2} gives the following relations:
%\begin{eqnarray}
%&&d_0^2=0,\quad d_0d_1-d_0d_1=0, \quad d_1d_1+d_0d_2+d_2d_0=0,\nonumber\\ 
%&&d_1d_2-d_2d_1=0, \quad d_2d_2=0,
%\end{eqnarray}
%which are in agreement with \rf{rel}.\\
%%%%%%%%%%%%%%%%%%%%%%%%%%%%%%%%%%%%%%%%%%%%%%%%%%%%%%%%%%%%%%%%%%%%%%%%%%%%%%%%

\noindent {\bf 2.2. Yang-Mills equations via generalized Maurer-Cartan equation.} \\
It is easy to see that the generalized Maurer-Cartan (GMC) equation for the YM $L_{\infty}$ algebra gives us the Yang-Mills equations. The GMC equation is:
\begin{eqnarray}\label{mc}
\mQ\mA+\frac{1}{2!}[\mA,\mA]_h+\frac{1}{3!}[\mA,\mA,\mA]_h=0,
\end{eqnarray}
where $\mA\in \mF^1_{\mathfrak{g}}$. Then
\begin{eqnarray}
&&\mQ\mA=\m \mA, \quad [\mA,\mA]_h=2(\ud *(\mA\wedge\mA)+[\mA\wedge *\ud \mA]),\nonumber\\
&&[\mA,\mA,\mA]_h=[\mA\wedge*(\mA\wedge\mA)].
\end{eqnarray}
Therefore, \rf{mc} is equivalent to the YM equations:
\begin{eqnarray}
\ud*\mathbf{F}+[{\mA},*\mathbf{F}]=0, \quad \mathbf{F}=\ud \mA+\mA\wedge \mA.
\end{eqnarray}
Moreover, the infinitesimal symmetries of the GMC equation
\begin{eqnarray}\label{mcgt}
\mA \to \mA+\epsilon(\mQ u +[\mA,u]_h),
\end{eqnarray}
where $u\in \mF^0_{\mathfrak{g}}$ and $\epsilon$ is infinitesimal parameter, 
coincide with the infinitesimal gauge symmetries of the YM equations
\begin{eqnarray}\label{gt}
\mA\to \mA+\epsilon(\ud u +[\mA,u]).
\end{eqnarray}
However, the elements of the complex $\mF^{\bf{\cdot}}_{\mathfrak{g}}$ are global forms on a manifold M, while $\mA$ from the YM equations is a connection 1-form. This means that for a given open cover $U_{\{\alpha\}}$ of the manifold $M$, $\mA$ transforms by means of the exponentiated version of \rf{gt}
from one open neighborhood $U_{\alpha}$ to 
another one $U_{\beta}$ 
(which have nonempty intersection):
\begin{eqnarray}
\mA_{\beta}=s_{\beta\alpha}\mA_{\alpha}(s_{\beta\alpha})^{-1}+s_{\beta\alpha}\ud (s_{\beta\alpha})^{-1},
\end{eqnarray}
where $\mA_{\alpha}$ stands for the restriction of $\mA$ to $U_{\alpha}$ and 
$\{s_{\beta\alpha}\}$ is a set of  $G$-valued functions (G is a Lie group associated with $\mg$) on $M$, such that $s_{\alpha\alpha}=Id$. These group-valued functions should obey the following cocycle property, if there is a nonempty  intersection of three open sets:
\begin{eqnarray}\label{cle}
s_{\alpha\beta}s_{\beta\gamma}=s_{\alpha\gamma}.
\end{eqnarray}
So, these are complete data that determine a connection 1-form: the set $\mA_{\{\alpha\}}$ 
and a set of gluing functions $s_{\alpha\beta}$. We will call the connection 1-forms satisfying YM equations  as the {\it YM connections} .

In order to modify the "formal" correspondence between the YM and GMC equations, in such a way that it becomes  equivalence, one should consider a sheaf of YM $L_{\infty}$ algebras. 

Namely, one has to consider  
the sheaf of sections of the bundles of 0- and 1-forms. 
Then, on each open subset $U$ we will have the complex 
$\mF^{\bf{\cdot}}_{\mathfrak{g}, U}$, the elements of which are the sections of the sheaf of Lie algebra-valued functions and 1-forms on $U$. Hence, reproducing a YM $L_{\infty}$ algebra on each $\mF^{\bf{\cdot}}_{\mathfrak{g}, U}$ , we can construct a sheaf of YM $L_{\infty}$ algebras. 

For each open cover $U_{\{\alpha\}}$ one has a possibility to construct a global Maurer-Cartan equation associated with 
the sheaf of $L_{\infty}$ algebras (see e.g. \cite{hinich} for more precise treatment). However, the way, how Maurer-Cartan element should transform  
from its restriction to $U_{\alpha}$ to its restriction to $U_{\beta}$, is infinitesimally determined by means of the transformation \rf{mcgt} in order to make Maurer-Cartan equation  hold on the intersection. In general, the cocycle property \rf{cle} for the gluing functions of the Maurer-Cartan element is modified (infinitesimally) by means of $Q$-exact terms, but since in $\mF^{\bf{\cdot}}_{\mathfrak{g}}$ we have no elements of negative grading, \rf{cle} holds without any modification. Therefore we have the following statement.\\

\noindent{\bf Proposition 2.2.} {\it There is one-to-one correspondence between the solutions of the GMC equation associated with the sheaf of YM $L_{\infty}$ algebras on a given Riemannian manifold and the YM connections.}\\

\noindent {\bf Remark.} In principle, one could extend the complex $\mF^{\bf{\cdot}}_{\mathfrak{g}}$ by means of $\mF^{-1}_{\mathfrak{g}}=\mathfrak{z}$, which is the center of the reductive Lie algebra $\mathfrak{g}$.  In this case, we  get precisely "symmetry of symmetries" and the cocycle condition \rf{cle} will be modified as follows: $s_{\alpha\beta}s_{\beta\gamma}=s_{\alpha\gamma}t_{\alpha\beta\gamma}$,  where $t_{\alpha\beta\gamma}$ is a constant element corresponding to the center $\mathfrak{z}$, giving a twisted bundle. \\

\noindent{\bf 2.3. Detour Complex.} In previous subsection, we considered forms with values in a trivial vector bundle corresponding to adjoint representation, that is why 
we needed to consider a sheaf of $L_{\infty}$-algebras in order to reproduce the Maurer-Cartan element associated to 
the YM connections. However, for a given YM connection $\mA_0$ one can construct a "global"  
$L_{\infty}$ algebra on the complex $(\mF^{\bf{\cdot}}_{\mathfrak{g}},\mQ)$, where the forms take their values 
in the adjoint vector bundle with transition functions $s_{\alpha\beta}$. However, the differential in the complex should be deformed, i.e. it has the following form:
\begin{eqnarray}\label{detour}
&&0\xrightarrow{ }\Omega^{0}(M,E^{ad})\xrightarrow{\ud_{\mA_0}}\Omega^{1}(M,E^{ad})
\xrightarrow{\mathfrak{m}_{\mA_0} }\nonumber\\
&&\Omega^{D-1}(M,E^{ad})\xrightarrow{\ud_{\mA_0}}\Omega^{D}(M,E^{ad})\to 0,
\end{eqnarray}
where $\Omega^{k}(M,E^{ad})$ stands for $k$-forms with values in the adjoint bundle, and 
the operator $\mathfrak{m}_{\mA_0}$ is defined as follows:
\begin{eqnarray}
\mathfrak{m}_{\mA_0} \mathbf{\mA}=\ud_{\mA_0}*\ud_{\mA_0} \mathbf{A}+[\mathbf{A},*\mathbf{F}_0],
\end{eqnarray}
where $\mathbf{F}_0=\ud\mA_0+\mA_0\wedge\mA_0$ and $\mA\in \Omega^{1}(M,E^{ad})$. One can easily get this complex locally deforming  the $\mQ$-operator of the original complex $(\mF^{\bf{\cdot}}_{\mathfrak{g}},\mQ)$ in the following way:
\begin{eqnarray}
\mQ_{\mA_0}\cdot=\mQ\cdot+[\mA_0,\ \cdot]_h+\frac{1}{2}[\mA_0,\mA_0,\ \cdot]_h.
\end{eqnarray}
This is a consequence of the general statement about $L_{\infty}$-algebras: in physics literature it was  considered by Zwiebach \cite{zwiebach}, when he studied background independence. Namely, for a given solution of a Maurer-Cartan equation, one can deform the differential and all the n-linear operation by means of it in such a way that they will again satisfy the $L_{\infty}$ algebra. The deformation of the bilinear operation (the trilinear is left unchanged) in our case is given by 
 \begin{eqnarray}
[\mathbf{A},\mathbf{B}]_{\mA_0}= [\mathbf{\mA},\mathbf{\mB}]_h+[ \mA_0,\mathbf{A},\mathbf{B}]_h.
 \end{eqnarray}
 This shift just corresponds to the substitution of the exterior derivatives by the exterior covariant ones in 
 the  definition of the bilinear operation.
The complex \rf{detour} is known as "Detour Complex" and it drew the attention in both mathematics 
and  physics \cite{detour}.\\

\noindent{\bf 2.4. Multilinear forms and YM action.} Here we proceed through the same steps as we did in \cite{bvym} in order to express the Yang-Mills action on a Riemannian manifold as a Homotopy Chern-Simons (HCS) action. Throughout this subsection we assume the manifold M to be compact in order to be  able to integrate safely over $M$.  

First of all, we introduce a pairing on a complex $\mF^{\bf{\cdot}}_{\mathfrak{g}}$. Namely, for any 
given elements $f_1, f_2\in \mF^{\bf{\cdot}}_{\mathfrak{g}}$ the inner product is defined by the following formula:
\begin{eqnarray}
\langle f_1, f_2 \rangle=\int_M Tr(f_1\wedge f_2).
\end{eqnarray}

\noindent This inner product appears to be invariant under the action of the differential $\mQ$.\\

\noindent{\bf Proposition 2.3.} {\it Let $f_1, f_2\in \mF_{\mg}$. Then, the following relation holds:
\begin{eqnarray}
\langle \mQ f_1, f_2\rangle+(-1)^{n_{f_1}n_{f_2}}\langle \mQ f_2, f_1\rangle=0,
\end{eqnarray} 
where numbers $n_{f_1}, n_{f_2}$ represent the grading of appropriate elements.}\\ 

\noindent The proof can be obtained by straightforward calculation.

\noindent Now, we will define the following multilinear forms on the complex $\mF_{\mg}$, which appear to be very 
useful for the construction of the Yang-Mills action.

For any $f_1, f_2, f_3, f_4\in \mF_{\mg}$, one can define n-linear 
forms (n=2,3,4)
\begin{equation}
\{\cdot,..., \cdot\}_h: \mF_{\mg}\otimes ...\otimes \mF_{\mg}\to \mathbb{C} 
\end{equation}
in the following way:
\begin{eqnarray}
&&\{f_1,f_2\}_h=\langle \mQ f_1, f_2\rangle, \quad \{f_1,f_2,f_3\}_h=\langle [f_1, f_2]_h,f_3\rangle,\nonumber\\
&&\{f_1,f_2,f_3, f_4\}_h=\langle [f_1, f_2,f_3]_h,f_4\rangle.
\end{eqnarray}

\noindent One can show that these multilinear forms enjoy the following important property.

\vspace{3mm}

\noindent {\bf Proposition 2.4.}{\it The multilinear products, introduced in the definition above, are graded antisymmetric, i.e.
\begin{eqnarray}
\{f_1,...,f_i,f_{i+1},..., f_n\}_h=-(-1)^{n_{f_i}n_{f_{i+1}}}\{f_1,...,f_{i+1},f_{i},..., f_n\}_h.
\end{eqnarray}
}
The proof can be obtained by the direct calculation, using the Jacobi identity for $\mg$ and the properties of the Hodge star.  

Now we are ready to formulate the next Proposition.\\

\noindent{\bf Proposition 2.5.} {\it The Yang-Mills action 
\begin{eqnarray}
S_{YM}=\frac{1}{2}Tr\int_M(\mathbf{F}\wedge *\mathbf{F}), \quad \mathbf{F}=\ud \mA+\mA\wedge \mA
\end{eqnarray}
can be expressed as follows:
\begin{eqnarray}\label{cs}
S_{YM}=-\frac{1}{2}\langle \mQ\phi_{\mA},\phi_{\mA}\rangle-\frac{1}{6}\{\phi_{\mA},\phi_{\mA},\phi_{\mA}\}_h-\frac{1}{24}
\{\phi_{\mA},\phi_{\mA},\phi_{\mA}, \phi_{\mA}\}_h,
\end{eqnarray}
where $\phi_{\mA}$ stands for the Maurer-Cartan element in a sheaf of $L_{\infty}$ algebras, associated with a connection 1-form $\mA$. }\\

\noindent The $L_{\infty}$ algebra we have considered in this section is the underlying algebraic structure for the BV Yang-Mills action. In order to obtain it,
 one has to consider a Grassmann algebra with integer grading, then to tensor it with complex $(\mF^{\cdot},\mQ)$. Then the Maurer-Cartan element (the element of 
overall grading 1) can be written as follows: $\Phi=\omega+\mA+\mA^*+\omega^*$,  where $\omega,\mA,\mA^*,\omega^*$ have grading $1, 0, -1, -2$ 
 w.r.t. the grading in the Grassmann algebra, and they 
are $0-, 1-, (D-1)-, D-$ forms correspondingly. Inserting $\Phi$ into the homotopy Chern-Simons action \rf{cs}, one obtains the BV YM theory:
\begin{eqnarray}
S_{BVYM}=S_{YM}+\int_M(\mA^*\wedge \ud_{\mA}\omega+[\omega,\omega]\omega^*).
\end{eqnarray}
The pairing $\langle \cdot,\cdot\rangle$ gives the usual BV symplectic form: $\Omega_BV=\int (\delta\mA\wedge\delta\mA^*+\delta\omega\wedge\delta\omega^*)$.

\section{Yang-Mills $C_{\infty}$ algebra and CFT}

{\bf 3.1. $C_{\infty}$ and $A_{\infty}$ algebra in gauge theory.} 
The $L_{\infty}$ algebra, which we have considered in previous section possesses the underlying 
structure, which appears to be more fundamental. 

Again, we consider the Maxwell complex \rf{maxwell} of differential forms. 
We claim that this complex has the structure of homotopy commutative associative algebra on it. 
Let's introduce the corresponding multilinear operations:
\begin{eqnarray}
&&(\cdot, \cdot)_h: \mathcal{F}^i\otimes \mathcal{F}^j\to \mathcal{F}^{i+j},\nonumber\\
&&(\cdot, \cdot, \cdot)_h: \mathcal{F}^i\otimes \mathcal{F}^j\otimes \mathcal{F}^k\to 
\mathcal{F}^{i+j+k-1}.
\end{eqnarray}
The bilinear operation is defined by means of the following table:\\

$(f_1,f_2)_h$=
\begin{tabular}{|l|c|c|c|r|}
\hline
\backslashbox{$ f_2$}{$f_1$}&\makebox{$v$} & \makebox{$\mA$} & \makebox{$\mV$} & \makebox{$a$} \\
\hline
$w$ & $vw$& $\mA w$& $\mV w$&$a w$\\
\hline
$\mB$ &$v\mB$  & $(\mA,\mB)$   &   $\mB\wedge\mV$&0\\
\hline
$\mW$ & $v\mW$ &  $ \mA\wedge\mW$ &    0&0\\
\hline
$b$ & $vb$ &   0 &    0&0\\
\hline
\end{tabular}\\

\noindent where $f_1$ takes values in the set $\{v, \mA, \mV, a\}$ from the first row, $f_2$ takes values in the set $\{w, \mB, \mW, b\}$. Other elements in the table represent the value of bilinear operation 
$(f_1,f_2)_h$ for appropriately chosen $f_1$ and $f_2$. In the table above $v,w\in\mathcal{F}^0$; 
$\mA,\mB\in \mathcal{F}^1$; $\mV,\mW\in\mathcal{F}^2$; $a,b\in\mathcal{F}^3$. 
The bilinear operation $(\mA,\mB)$ is defined as follows:
\begin{eqnarray}
&&(\mA,\mB)=(\mA\wedge*\ud\mB)-(\mB\wedge*\ud\mA)+\ud *(\mA\wedge \mB).
\end{eqnarray}
The operation $(\ \cdot, \ \cdot, \ \cdot)_h$ is defined to be nonzero only when all the arguments belong to $\mF^1$. For $\mA$, 
$\mB$, $\mC$$\in \mF^1$ we have: 
\begin{eqnarray}
(\mA,\mB ,\mC )_h=\mA\wedge*(\mB\wedge\mC)-\mC\wedge*(\mA\wedge\mB).
\end{eqnarray}
By direct calculations given in Appendix A, one can show that these operations satisfy the following relations, which provide the structure of $C_{\infty}$ algebra. \\

\noindent{\bf Proposition 3.1.} {\it Let $a_1,a_2, a_3, a_4,b, c$ $\in$ $\mF_{\mg}$. Then the following relations hold:
\begin{eqnarray}
&&\mQ(a_1,a_2)_h=(\mQ a_1,a_2)_h+(-1)^{n_{a_1}}(a_1,\mQ a_2)_h,\nonumber\\
&&(a_1,a_2)_h=(-1)^{n_{a_1}n_{a_2}}(a_2,a_1)_h,\nonumber\\
&&\mQ(a_1,a_2, a_3)_h+(\mQ a_1,a_2, a_3)_h+(-1)^{n_{a_1}}(a_1,\mQ a_2, a_3)_h+\nonumber\\
&&(-1)^{n_{a_1}+n_{a_2}}( a_1, a_2, \mQ a_3)_h=((a_1,a_2)_h,a_3)_h-(a_1,(a_2,a_3)_h)_h
,\nonumber\\
&&(-1)^{n_{a_1}}(a_1,(a_2,a_3,a_4)_h)_h+((a_1,a_2,a_3)_h,a_4)_h=\nonumber\\
&&((a_1,a_2)_h,a_3,a_4)_h-(a_1,(a_2,a_3)_h,a_4)_h+\nonumber\\
&&(a_1,a_2,(a_3,a_4)_h)_h,
\nonumber\\
&&((a_1,a_2, a_3)_h,b,c)_h=0.
\end{eqnarray} }

\noindent If we tensor complex $(\mF^{\bf\cdot},\mQ)$ with a universal enveloping algebra for a Lie algebra $\mg$, one obtains that inherited operations $(\ \cdot,\ \cdot)_h$ and $(\ \cdot,\ \cdot, \ \cdot)_h$ satisfy the relations of  $A_{\infty}$ algebra. Then, considering appropriate antisymmetrization like in \cite{ladamarkl},
 one obtains a YM homotopy Lie algebra, e.g. it is easy to see that
\begin{eqnarray}
 [a,b]_h=(a,b)_h-(-1)^{n_a n_b}(b,a)_h.
 \end{eqnarray}
One can write down the Maurer-Cartan equation and its symmetries also in the case of $A_{\infty}$ algebra:
\begin{eqnarray}
\mQ\mA+(\mA,\mA)_h+(\mA,\mA,\mA)_h=0\nonumber\\
\mA\to \mA+\epsilon (\mQ u +(u,\mA)_h-(\mA,u)_h),
\end{eqnarray}
which also locally gives a Yang-Mills equation. One can also define the sheaf of $A_{\infty}$ algebras as we did in the previous section for the $L_{\infty}$ case and make identification between solutions of the GMC equation for YM $A_{\infty}$ algebra defined above and YM connections.\\

\noindent {\bf 3.2. Open string CFT and Maxwell complex.} 
We consider the open String Theory in D-dimensional space on the upper half-plane, and we fix the 
operator products between the coordinate fields as follows (see e.g. \cite{pol}): 
\begin{eqnarray}\label{ope}
X^{\mu}(z_1)X^{\nu}(z_2)\sim -\alpha'\eta^{\mu\nu}\log|z_1-z_2|^2-\alpha'\eta^{\mu\nu}\log|z_1-\b z_2|^2,
\end{eqnarray}
where $\eta^{\mu\nu}$ is the constant metric in the flat $D$-dimensional space of either Euclidean or Minkowski signature. The associated BRST operator has the following form, see e.g. \cite{pol}:
\begin{eqnarray}\label{brst}
Q=\oint dz (cT+bc\p c +\frac{3}{2}\p^2 c) , \quad T=-(2\alpha')^{-1}\p X^{\mu}\bp X^{\mu},
\end{eqnarray}
where normal ordering is implicit,  $b$ and $c$ are the usual ghost fields of conformal weights $2$ and $-1$ correspondingly with the 
operator product 
\begin{eqnarray}
c(z)b(w)\sim \frac{1}{z-w}. 
\end{eqnarray} 
We define the ghost number operator $N_g$ by 
\begin{eqnarray}
N_g=3/2+1/2(c_0b_0-b_0c_0)+\sum^{\infty}_{n=1}(c_{-n}b_n-b_{-n}c_n). 
\end{eqnarray} 
The constant shift (+3/2) is included to make the ghost number of the $SL(2,\mathbb{C})$-invariant vacuum state $|0\rangle$ be 
equal to 0. We denote the ghost number of a state V as $|V|$. \\
Firstly, we  will consider the states corresponding to the operators
\begin{eqnarray}\label{f0}
&&\rho_u=u(X), \quad \phi_{\mA}=cA_{\mu}(X)\p X^{\mu}-\alpha'\p c\p_{\mu}A^{\mu}(X),\nonumber\\ 
&&\psi_{\mW}=\alpha'c\p c W_{\mu}(X)\p X^{\mu}, \quad \chi_{a}=\alpha'^2 c\p c \p^2 c a(X),
\end{eqnarray}
 associated with functions $u(x)$, $a(x)$ and 1-forms 
$\ \mathbf{A}=A_{\mu}(x)dx^{\mu}$, $\ \mathbf{W}=$ $W_{\mu}(x)dx^{\mu}$.  
The resulting space $\tilde \mF$, spanned by the states like those associated with \rf{f0}, is invariant under the action of the BRST operator. Moreover, we see that if we denote by $\tilde \mF^{i}$ the subspace spanned by the states of ghost number $i$, one obtains the flat space version of the complex \rf{maxwell}. Really, the action of $Q$ operator is given by:
\begin{eqnarray}\label{qact}
Q\rho_{u}=2\phi_{\ud u},\quad Q\phi_{\mathbf{A}}=2\psi_{\mathfrak{m} \mA},\quad Q\psi_{\mathbf{W}}=-\chi_{\mathbf{div} \mW}, \quad 
Q\chi_{a}=0,
\end{eqnarray}
where $\mathfrak{m}\mA=(\p_{\mu}\p^{\mu}A_{\nu}-\p_{\nu}\p_{\mu}A^{\mu})dx^{\nu}$ and 
$\mathbf{div} \mW=\p_{\mu}W^{\mu}$. Then, identifying the space of 1-forms (of ghost number 2) with 
$(D-1)$-forms and $0$-forms (of ghost number 3) with $D$-forms by means of the Hodge star, we get the complex $(\mF^{\bf\cdot},Q)$; moreover, the ghost number is identified with the grading in the Maxwell complex. 
One can see that it is natural to extend the Maxwell complex by means of the contractible complex, given by the states associated with the following operators, corresponding to 0-forms:
\begin{eqnarray}
\varphi_a=\alpha' \p ca(X), \quad \xi_b=Q\varphi_b=\alpha'c\p^2 cb(X)+\alpha'2c\p c\p_{\mu}b(X)\p X^{\mu}. 
\end{eqnarray}
One can easily find out that the resulting extended BRST subcomplex coincides with the extended Maxwell complex.\\

\noindent{\bf 3.3. Reminder of Lian-Zuckerman construction.} In this subsection, we give only the necessary facts about Lian-Zuckerman operations. For more information 
one should consult \cite{lz}. 

Let us consider some chiral algebra. Let $T(z)$ denote the appropriate Virasoro element (i.e. energy momentum tensor) and let $Q$ be an associated semi-infinite cohomology 
operator  (i.e. BRST operator):
\begin{eqnarray}
Q=\oint dz (cT+bc\p c+\frac{3}{2}\p^2 c). 
\end{eqnarray}
Then, one can define the following bilinear 
operation on the corresponding space of states:
\begin{eqnarray}
\mu(a_1,a_2)=Res_z\frac{a_1(z)a_2}{z},
\end{eqnarray}
where $a(z)$ is the vertex operator associated with the state $a$. Lian and Zukerman have showed 
that this bilinear operation is homotopy commutative and associative w.r.t. the operator $Q$, namely the following Proposition hold.\\

\noindent{\bf Proposition 3.2.} \cite{lz}
{\it The bilinear operation $\mu$ satisfies the following relations:
\begin{eqnarray}\label{lzrel}
&&\mQ\mu(a_1,a_2)=\mu(\mQ a_1,a_2)+(-1)^{|a_1|}\mu(a_1,\mQ a_2),\nonumber\\
&&\mu(a_1,a_2)-(-1)^{|a_1||a_2|}\mu(a_2,a_1)=\nonumber\\
&&Qm(a_1,a_2)+m(Qa_1,a_2)+(-1)^{|a_1|}m(a_1,Qa_2),\nonumber\\
&&\mu(\mu(a_1,a_2),a_3)-\mu(a_1,\mu(a_2,a_3))=\nonumber\\
&&Qn(a_1,a_2,a_3)+n(Qa_1,a_2,a_3)+(-1)^{|a_1|}n(a_1,Qa_2,a_3)+\nonumber\\
&&(-1)^{|a_1|+|a_2|}n(a_1,a_2,Qa_3),
\end{eqnarray}
where 
\begin{eqnarray}
&&m(a_1,a_2)=\sum_{i\ge 0}\frac{(-1)^i}{i+1}Res_wRes_{z-w}(z-w)^iw^{-i-1}b_{-1}\nonumber\\
&&\qquad\qquad\quad (a_1(z-w)a_2)(w)\mathbf{1},
\nonumber\\
&&n(a_1,a_2,a_3)=\sum_{i\ge 0}\frac{1}{i+1}Res_zRes_w w^iz^{-i-1}(b_{-1}a_1)(z)a_2(w)a_3+\nonumber\\
&&(-1)^{|a_1||a_2|}\sum_{i\ge 0}\frac{1}{i+1}Res_wRes_z z^iw^{-i-1}(b_{-1}a_2)(w)a_1(z)a_3.
\end{eqnarray}}

Thus, on the level of cohomology w.r.t. the operator $Q$, this algebra turns out to be commutative and associative. This fact was very useful for the study of the algebra of states in string backgrounds \cite{lz}. \\

\noindent{\bf 3.4. Open string CFT and Lian-Zuckerman products via point-split\-ting.}  
The homotopy commutative associative algebra which we considered in subsection 3.3, is related to chiral algebra. However, in the case of open string CFT, the operator products involve logarithms, and in such a way the Lian-Zuckerman operation should be modified appropriately. At the same time, since we want to study only the lowest order $\alpha'$-corrections, we can throw away the logarithms which are always accompanied with $\alpha'$. So, let us be more concrete. 
In this section, all operators are assumed to have position on the real line.  If we take two 
operators $V(t_1)$ and $V(t_2)$, where $t_1,t_2\in \mathbb{R}$, their operator product has the following form:
\begin{eqnarray}\label{opestr}
V(t_1)W(t_2)= \sum_{k,l}(t_1-t_2)^{-l}(V,W)_l^{(k)}(t_2)(\alpha'\log|(t_1-t_2)/\mu|^2)^k.
\end{eqnarray}
The logarithms in OPE create higher $\alpha'$ corrections, so we will omit them in our considerations 
since we will be interested in the lowest orders in $\alpha'$; instead of the full OPE \rf{opestr} we consider the reduced OPE
 \begin{eqnarray}\label{reduced}
(V(t_1)W(t_2))_R= \sum_{l}(t_1-t_2)^{-l}(V,W)_l^{(0)}(t_2).
\end{eqnarray}
To simplify the notation, we will omit the letter $R$ and superscript $(0)$ in the following. 
Now we will reformulate the Lian-Zuckerman operations from the subsection 3.3. via point-splitting regularization.\\
For open string states $V$ and $W$, we define a bilinear operation $\t \mu(V,W)$ such that the operator associated with the state is given by 
\begin{eqnarray}
{\t \mu}(V,W)(t)=\mathcal{P}_0V(t+\epsilon)W(t),
\end{eqnarray}
where $\epsilon$ is a real parameter and $\mathcal{P}_0$ is a projection on the $\epsilon$-independent term. 
It is clear from this definition, that the BRST operator acts as a derivation for $\mu$, and for the chiral algebra it coincides with the Lian-Zuckerman bilinear operation, considered in subsection 3.3..

Our next aim is to define the analogue of the operations $n,m$ from the previous subsection via point-splitting language. 
Before that, we need to give some prerequisite notation. Let $V(t)$ and  $W(t)$ be the operators associated with the
state $V$ and $W$ correspondingly. We define a "naively integrated"
operator $L^{-1}_{-1}V(t)$  in
the following way:
\begin{eqnarray}
&&L^{-1}_{-1}V(t_1)W(t_2)\equiv\nonumber\\
&&\sum_{l\neq 1}\frac{1}{1-l}(t_1-t_2)^{-l+1}}{(V,W)_l^{(0)}(t_2)+(V,W)_{1}^{(0)}(t_2)\log(t_1-t_2),
\end{eqnarray}
which, in the case of chiral algebra, corresponds to the indefinite integration of the OPE.
We need to note that this is just a notation, since the operator $L^{-1}_{-1}V(t_1)$ in most of cases  
does not correspond to any state. Moreover, in the definition above we neglected the integration of logarithms 
which are unnecessary in the following since (as it was already noted) they produce higher order $\alpha'$-corrections. 
Let us make one more notation:
\begin{eqnarray}
(V,W)^{\epsilon,s}(t)\equiv
\sum^{\infty}_{l=1}\epsilon^{-l}(V,W)_l^{(0)}(t),
\end{eqnarray}
i.e. $(V,W)^{\epsilon,s}(t)$ is the singular part of the operator product $V(t+\epsilon)W(t)$. 
Now we are ready to give the expression for the operation $m$ in $\epsilon$-language. 
For open string states $V$ and $W$ we define ${\t m}(V,W)$ in 
such a way, that the operator corresponding to the state ${\t m}(V,W)$ is given by 
\begin{eqnarray}
{\t  m}(V,W)(t)=-\mathcal{P}_0L^{-1}_{-1}b_{-1}(V,W)^{\epsilon,s}(t-\epsilon).
\end{eqnarray}
One can show that in the case of vertex algebra, $\t m\equiv m$. If $V,W$ are the states in the chiral algebra, then  
\begin{eqnarray}
L_{-1}m(V,W)(t)=\sum^n_{i\ge 0}\frac{(-1)^i}{(i+1)!}b_{-1}\p^{i+1}(V,W)^{0}_{i+1}(t).
\end{eqnarray}
At the same time,
\begin{eqnarray}
(V,W)^{\epsilon,s}(t-\epsilon)=\sum^n_{i\ge 0}\epsilon^{-i-1}(V,W)^{0}_i(t-\epsilon).
\end{eqnarray}
Therefore, $m(V,W)=\t m(V,W)$, when $V,W$ belong to some vertex algebra. 
The expression for a trilinear operation in $\epsilon$-language is given as follows. 
For open string states $U$,$V$,$W$, a trilinear operation 
${\t n}(U,V,W)$ is defined in 
such a way, that the operator corresponding to the state ${\t n}(U,V,W)$ is given by 
\begin{eqnarray}
&&{\t  n}(U,V,W)(t)=\mathcal{P}_0(L^{-1}_{-1}b_{-1}U)(t+\epsilon)(V,W)^{\epsilon,s}(t)+\nonumber\\
&&(-1)^{|U||V|}\mathcal{P}_0(L^{-1}_{-1}b_{-1}V(t+\epsilon))(U,W)^{\epsilon,s}(t).
\end{eqnarray}
We show below, that in the case of vertex algebra it coincides with the operation $n$. 
Let $U$, $V$, $W$ be the states in some chiral algebra. Then
\begin{eqnarray}
&&n(U,V,W)=\sum_{i\ge 0}\frac{1}{i+1}(b_{-1}U,(V,W)_{i+1})_{-i}+\nonumber\\
&&(-1)^{|U||V|}\frac{1}{i+1}(b_{-1}U,(V,W)_{i+1})_{-i},
\end{eqnarray}
on the other hand, 
\begin{eqnarray}
\t n(U,V,W)=\mathcal{P}_0\sum_{k\ge0}\frac{1}{k+1}\epsilon^{k+1}(U,(V,W)^{\epsilon,s})_{-k}(t)+\nonumber\\
(-1)^{|U||V|}\mathcal{P}_0\sum_{k\ge0}\frac{1}{k+1}\epsilon^{k+1}(V,(U,W)^{\epsilon,s})_{-k}(t).
\end{eqnarray}
Comparing the powers of $\epsilon$, we find out that $m(U,V,W)=\t m(U,V,W)$. In the following, we therefore omit tildes for operations  $\t \mu, \t m, \t n$.
  
We believe that $m,n$ are first two operations in the conjectured homotopy algebra of the perturbation theory for the open string sigma models regularized via point-splitting \cite{zeit3}. In the next subsection, we will see evidence of this statement, when we obtain the the YM $C_{\infty}$ algebra from the Lian-Zuckerman operations.

One has to mention that the definitions we gave should be changed, when  the contribution of the logarithms are taken into account, since 
the operations we introduced, most likely will not satisfy the relations \rf{lzrel} if  the $\alpha'$-corrections of higher degree are included. We postpone 
the study of this question for subsequent publications. 

We note that in \cite{ym}-\cite{zeit3} we considered the bilinear operation of another type: 
\begin{eqnarray}
R(U,V)(t)=\mathcal{P}_0[U(t+\epsilon),V(t)]-(-1)^{|U| |V|}
\mathcal{P}_0[V(t+\epsilon),U(t)]
\end{eqnarray}
on the tensor product of the space of operators of the open string and some Lie algebra $\mg$. This operation (when reduced to the operators of the Maxwell complex) 
led to the $L_{\infty}$ algebra in YM theory. \\

Before we proceed to the explicit calculations, let's stop for a moment to give a pure algebraic sense to the introduction of the $\epsilon$-parameter. Since all the operations have the projection operator 
$\mathcal{P}_0$ in their definition, it would be natural to throw it away and to see what happens in this case. 
Let us consider the following operation:
\begin{eqnarray}
\mu^{(i)}(V,W)\equiv Res_z z^{-i-1}V(z)W 
\end{eqnarray}
for any $i\in \mathbb{Z}_{\ge 0}$, such that $\mu\equiv \mu^{(0)}$. Let's denote the regular part of the OPE as follows:
\begin{eqnarray}
\mu^{\epsilon}(V,W)(t)\equiv V(t+\epsilon)W(t)-(V,W)^{\epsilon,s}=\sum_{k\ge 0} \epsilon^{k}\mu^{(k)}(V,W)(t).
\end{eqnarray}
One can check that if $U$, $V$, $W$ are states in some chiral algebra, then 
\begin{eqnarray}
&&\mu^{(l)}(\mu^{(k)}(U,V)W)-\mu^{(k)}(U(\mu^{(l)}(V,W))=\nonumber\\
&&Qn^{(k,l)}(U,V,W)+n^{(k,l)}(QU,V,W)+(-1)^{|V|}n^{(k,l)}(U,QV,W)+\nonumber\\
&&(-1)^{|U|+|V|}n^{(k,l)}(U,V,QW)
\end{eqnarray}
for some operation $n^{(k,l)}$. Therefore, 
\begin{eqnarray}
&&\mu^{\epsilon}(\mu^{\epsilon}(U,V)W)-\mu^{\epsilon}(U(\mu^{\epsilon}(V,W))=\nonumber\\
&&Qn^{\epsilon}(U,V,W)+n^{\epsilon}(QU,V,W)+(-1)^{|U|}n^{\epsilon}(U,QV,W)+\nonumber\\
&&(-1)^{|U|+|V|}n^{\epsilon}(U,V,QW),
\end{eqnarray}
where $n^{\epsilon}(U,V,W)=\sum_{k,l\ge 0}\epsilon^{-k-l} n^{(k,l)}(U,V,W)$. Therefore, we have an extension of the Lian-Zuckerman homotopy algebra. We will return to this and other related questions in 
\cite{zeitnew}.\\

\noindent{\bf 3.5. From CFT of open string to $C_{\infty}$ algebra of YM theory.}
In this subsection, we will finally find out how Lian-Zuckerman products are related to YM $C_{\infty}$ algebra. 
Let's consider the embedding of the (extended) Maxwell complex in the BRST complex of open string from  subsection 3.3. 
Calculating the OPE carefully, one can get the lowest order $\alpha'$ corrections to the $\mu$-operation on the extended Yang-Mills complex. In such a way 
one gets an operation which we will denote by the same letter $\mu$ acting in the following way:
\begin{eqnarray}
\mu(\cdot, \cdot): \mathcal{F}_{ext}^i\otimes \mathcal{F}_{ext}^j\to \mathcal{F}_{ext}^{i+j}.
\end{eqnarray}
The explicit values of it are given in the following table (see Appendix B):\\

\begin{center}$\mu(f_1,f_2)$=
\end{center}

\noindent\begin{tabular}{|l|c|c|c|c|c|r|}
\hline
\backslashbox{$f_2$}{$f_1$}&\makebox{$v$}   & \makebox{$\mau$}     &\makebox{$\mA$}       &  \makebox{ $\ma$}    &\makebox{ $\mV$}  & \makebox{$a$} \\
\hline
$w$        & $vw$                                            &    $\mau w$                   & $\mA w-$                                        &     $w\ma-$                                         & $\mV w$    &$a w$\\
               &                                                        &                                         &$*(\ud w\wedge*\mA) $                  &   $*(\ud w\wedge*\ma)$               &                    &  \\
\hline
$\mv$ & $v\mv$                                             &        0                          &     $*\mA\mv$                                    &    $\ma\mv$                                   &      0          &0\\
\hline
$\mB$ &$v\mB +$                                        &      $-*\mB \mau$        & $(\mA,\mB) +$                                 &    $-\ud*(\mB\wedge*\ma)$      &  $\mB\wedge\mV$&0\\
             & $*(\ud v\wedge *\mB)$            &                                       &    $ (\mA\wedge*\mB)$                  &                                                       &                                  &   \\  
\hline
$\mb$ & $v\mb-$                                       &    $\mau\mb$              &$-\ud*(\mA\wedge*\mb)$               &    0                                                &  0                           & 0\\
            & $*(\ud v\wedge*\mb)$                &                                    &                                                             &                                                       &                               &\\
\hline
$\mW$ & $v\mW$                                            &      0                          &     $ \mA\wedge\mW$                     &     0                                                & 0&               0\\
\hline
$b$ & $vb$                                                      &           0                       &      0                                                   &         0                                          &   0 &                 0\\
\hline
\end{tabular}\\

\vspace{5mm}

\noindent Here $f_1,f_2$ take values in the sets $\{v,\mau,\mA,\ma,  \mV ,a\}$ and $\{v,\mv,\mB,\mb,  \mW ,b\}$ correspondingly. 
Other elements in the table correspond to the appropriate values of $\mu(f_1,f_2)$. In the table above 
$v,w\in \mF^0$; $\mau,\mv\in\mG^1$;  $\mA,\mB\in\mF^1$; $\ma,\mb\in\mG^2$; $\mW,\mV\in\mF^2$; $a,b\in\mF^3$. 

Next, we consider the lowest 
order $\alpha'$-corrections to  the operation $m$. The resulting operation which we will denote the same letter $m$, acts as follows on the extended Maxwell complex:
\begin{eqnarray}
m(\cdot,\cdot): \mathcal{F}_{ext}^i\otimes \mathcal{F}_{ext}^j\to \mathcal{F}_{ext}^{i+j-1}.
\end{eqnarray}
Explicit calculation (see Appendix B) shows that the value of $m$ is nonzero only if both arguments belong to $\mF^1$:
\begin{eqnarray}\label{mexpl}
m(\mA,\mB)=2*(\mA\wedge*\mB)
\end{eqnarray}
for  $\mA$,$\mB\in\mF^1$. As for the operation $n$, its lowest $\alpha'$ corrections produce the following operation on the $(\mF_{ext}^{\bf\cdot},\mQ)$:
\begin{eqnarray}
n(\cdot, \cdot, \cdot): \mathcal{F}_{ext}^i\otimes \mathcal{F}_{ext}^j\otimes \mathcal{F}_{ext}^k\to 
\mathcal{F}_{ext}^{i+j+k-1},
\end{eqnarray}
such that it is nonzero only if all the three arguments belong to $\mF^1$ or if 
one of the first two arguments belong to $\mG^2$, namely (see Appendix B):
\begin{eqnarray}\label{nexpl}
&&n(\mA,\mB,\mC)=*\mA\wedge*(\mB\wedge *\mC)-*\mC\wedge*(\mA\wedge*\mB)+\nonumber\\
&&\mA\wedge*(\mB\wedge\mC)-\mC\wedge*(\mA\wedge\mB),\nonumber\\
&&n(\ma,\mA,\mB)=n(\mA,\ma,\mB)=-2*(\mA\wedge*\mB)\wedge\ma,
\end{eqnarray}
where $\mA,\mB,\mC\in\mF^1$ and $\ma\in\mG^2$. As one could suspect, it appears that the induced 
operations $\mu, m,n$ on the complex  $(\mF^{\bf \cdot},\mQ)$ satisfy, precisely, the same relations 
as original ones on the chiral algebra.
\\

\noindent{\bf Proposition 3.3.} 
{\it The operations $\mu, n,m$ satisfy the following relations:
\begin{eqnarray}
&&\mQ\mu(a_1,a_2)=\mu(\mQ a_1,a_2)+(-1)^{n_{a_1}}\mu(a_1,\mQ a_2),\nonumber\\
&&\mu(a_1,a_2)-(-1)^{n_{a_1}n_{a_2}}\mu(a_2,a_1)=\nonumber\\
&&\mQ m(a_1,a_2)+m(\mQ a_1,a_2)+(-1)^{n_{a_1}}m(a_1,\mQ a_2),\nonumber\\
&&\mu(\mu(a_1,a_2),a_3)-\mu(a_1,\mu(a_2,a_3))=\nonumber\\
&&\mQ n(a_1,a_2,a_3)+n(\mQ a_1,a_2,a_3)+(-1)^{n_{a_1}}n(a_1,\mQ a_2,a_3)+\nonumber\\
&&(-1)^{n_{a_1}n_{a_2}}n(a_1,a_2,\mQ a_3),
\end{eqnarray}
where $a_i\in\mF^{\bf \cdot}_{ext}$.}\\

\noindent The proof of this statement is given in Appendix A. 

Now, from the explicit form of the operations  $\mu, n,m$ one can restore the YM $C_{\infty}$ algebra. Namely, we have the following Proposition.\\

\noindent{\bf Proposition 3.4.} {\it The operations
\begin{eqnarray}
&&\mu'(a_1,a_2)=\mu(a_1,a_2)-\frac{1}{2}\Big(\mQ m(a_1,a_2)+m(\mQ a_1,a_2)\nonumber\\
&&+(-1)^{n_{a_1}}m(a_1,\mQ a_2)\Big),\nonumber\\
&&n'(a_1,a_2,a_3)=n(a_1,a_2,a_3)+\frac{1}{2}\Big(\mu\big(m(a_1,a_2),a_3\big)\nonumber\\
&&-(-1)^{n_{a_1}}\mu\big(a_1,m(a_2,a_3)\big)+m(\mu(a_1,a_2),a_3)-m(a_1,\mu(a_2,a_3))\Big)
\end{eqnarray}
act invariantly on Maxwell complex and coincide with the operations $(\cdot,\cdot)_h$ and $(\cdot,\cdot,\cdot)_h$.}\\

\noindent In other words, the homotopy Abelian associative algebra of $\mu, n, m$ is {\it quasiisomorphic} to some other 
homotopy abelian associative algebra on 
$(\mF_{ext}^{\cdot},\mQ)$, which, being reduced 
to the complex $(\mF^{\cdot},\mQ)$, coincides with $C_{\infty}$ algebra of YM. 

Let's return to the algebraic structure generated on $\mF^{\bf{\cdot}}_{ext}$ by the operations $\mu,n,m$. As in the case of YM $C_{\infty}$ algebra, 
one can tensor $\mF^{\bf {\cdot}}_{ext}$ with some Lie algebra $\mg$ and obtain the $A_{\infty}$ algebra.  
Let's write down the corresponding Maurer-Cartan equation with symmetries for the corresponding 
$A_{\infty}$ algebra: 
\begin{eqnarray}
&&\mQ\Phi+\mu(\Phi,\Phi)+n(\Phi,\Phi,\Phi)=0,\nonumber\\
&&\Phi\to Qu+\mu(\Phi,u)-\mu(u,\Phi).
\end{eqnarray}
Here the Maurer-Cartan element is represented (since we are considering the extended complex) by means of the sum $\Phi=\mA+\mau$, where 
$\mA\in \mF^1$ and 
$\mau\in \mG^1$, while the symmetries are generated by $\lambda=u\in \mF^0$. It is easy to check that the Maurer-Cartan equations are equivalent to Yang-Mills equations for $\mA$ and the additional equation for $\mau$: 
\begin{eqnarray}
&&\ud_{\mA}*\bf{F}(\mA)=0, \quad \bf{F}(\mA)=\ud\mA+\mA\wedge\mA,\\
&&\label{ag}\mau=-*(\mA\wedge*\mA).
\end{eqnarray}
The symmetries coincide with usual gauge symmetries. The equation \rf{ag} has the interesting meaning, since the appearance of such term of the type 
$A_{\mu}A^{\mu}$ is provided by the presence of an additional {\it bivertex} operator in the point-splitting regularization of the open string sigma model.  The presence of such  bivertex operator was already noticed in the case of closed string sigma model (see \cite{zeit3}).\\

\noindent{\bf 3.6. Gauge fixing procedure.} In subsection 3.5 we considered the nontrivial extension of the YM $C_{\infty}$ algebra on the extended Maxwell complex, given by Lian-Zuckerman products. In this subsection, we concentrate on the trivial extension, namely we will study the complex $(\mF^{\cdot}_{ext},\mQ)$ with operations $(\cdot,\cdot)_h$ and
$(\cdot,\cdot,\cdot)_h$ such that these operations are equal to zero when one of the arguments belongs to contractible complex $\mG^i$. 

From  subsection 3.2. we know that there exists an embedding of the extended Maxwell complex on flat space in the BRST complex of open string. However, one can partly hold this embedding in the case of the manifold with nontrivial metric, namely one can embed the extended Maxwell complex in the following 
superspace:
\begin{eqnarray}
\mathfrak{S}=\big(\Omega^0(M)\oplus\Omega^1(M)\big)[c_0,c_1,c_{-1}],
\end{eqnarray}
where $c_0,c_1,c_{-1}$ are anticommuting variables (corresponding to the modes of $c$-ghost). The $\mQ$ operator is given by 
\begin{eqnarray}
\mQ=c_1\ud+c_{-1}*\ud*+c_0(\ud*\ud*+*\ud*\ud)-c_{-1}c_{1}\frac{\p }{\p c_{0}}.
\end{eqnarray}
The grading operator is given by $N=\sum c_i\frac{\p}{\p c_i}$. The embedding of subspaces $\mF^{i}$ is given by the following formulae:
\begin{eqnarray}
&&\mF^0\ni u\to u, \quad \mF^1\ni\mA\to c_1\mA+c_0*\ud*\mA, \nonumber\\
&&\mF^2\ni \mW\to c_0c_1*\mW, \quad \mF^3\ni a\to c_{-1}c_{0}c_{1}*a,\nonumber\\
&& \mG^1\ni \mau\to -c_0\mau, \quad \mG^2\ni\ma\to c_{-1}c_{1}*\ma+c_0c_1*\ud*\ma.
\end{eqnarray}
The invariant pairing on the extended Maxwell complex is given by the formula:
\begin{eqnarray}\label{pairing}
\langle\Phi,\Psi\rangle=\int_{M}\int d{c_{-1}}d{c_{0}}d{c_{1}}(\Phi\wedge*\Psi).
\end{eqnarray}
One can see that the embedded complex $(\mF^{\bf\cdot}_{ext},\mQ)$ is invariant under the action of the operator ${\bf b}=\frac{\p}{\p c_0}$ (it corresponds to   $b_0$ mode of  b-ghost).  This operator acts explicitly on $(\mF^{\bf\cdot}_{ext},\mQ)$ in following way:

\begin{displaymath}
\xymatrixcolsep{50pt}
\xymatrixrowsep{5pt}
\xymatrix{
\Omega^0 & \Omega^1 \ar[l]_{*\ud*}& \Omega^{D-1} \ar[l]_{*} \ar[ddddl] & \Omega^D \ar[l]_{*\ud*} \ar[ddddl]^{-Id}\\
&_{(-1)^{D-1}\ud*}&&\\
 & \bigoplus &\bigoplus & \\
\qquad &\qquad _{\ud*} &  & \qquad \\
\qquad  & \Omega^0 \ar[uuuul]^{Id(-1)^D} \ar[r]^{*\ud*\ud*} & \Omega^D \ar[uuuul] & \qquad
}
 \end{displaymath}

One can easily obtain that  the $\bf{b}$ operator is self-adjoint w.r.t. the pairing \rf{pairing}. Moreover, it has trivial cohomology, and  commutation relations between $\mQ$ and $\bf{b}$ operators are given by the simple formula: 
\begin{eqnarray}\label{qb}
[\mQ,{\bf{b}}]=\Delta, 
\end{eqnarray}
where $\Delta=*\ud*\ud+\ud*\ud*$ is the Laplacian. We have the following Proposition.\\

\noindent {\bf Proposition 3.5.} {\it The space $\mF=\oplus^3_{i=0}\mF^i_{ext}$ allows the Hodge-type decomposition:
\begin{eqnarray}\label{hodge}
\mF=\ker\Delta\oplus Im{\bf{b}}\oplus Im\mQ.
\end{eqnarray}}
It is clear that the subspace $Im{\bf{b}}$ is the Lagrangian submanifold in $\mF_{ext}$ w.r.t. the symplectic pairing given by \rf{pairing}. It is not hard to see that 
on the BV language the Lagrangian submanifold $Im{\bf{b}}$ corresponds to the choice of a Lorenz gauge in YM theory. Here, we wrote it down in the 
cohomological way (see \rf{qb},\rf{hodge}). In \cite{costello}, it was noted that the amount of BV theories which allow such type of gauge fixing is very limited. It was also said that it was not known whether gauge fixing of this type could be performed for usual YM theory (however, the author finds that this can be done for the 
first order formulation of YM theory, to which we will return in section 4). Here we fill this gap.\\  

\noindent {\bf 3.7. $G_{\infty}$ algebra in YM theory?} The origin of the operator $\bf{b}$, which, as we have already seen, gives the gauge fixing condition, lies in the conformal field theory. As we have already noted,
when one embeds the extended Maxwell complex in BRST complex, this operator corresponds to the $b_0$ mode of the $b$-ghost. It is known \cite{lz}, that the operation $\{\cdot, \cdot\}_h$ given by the formula
\begin{eqnarray}
\{U,V\}=Res_z(b_{-1}U)(z)V=b_0\mu(U,V)-\mu(b_0U,V)-(-1)^{|U|}\mu(U,b_0V),
\end{eqnarray}
where $U,V$ are the 
states in some chiral algebra, generates a homotopy Gerstenhaber algebra on the space of states of chiral algebra.  
One can expect that the induced operation 
\begin{eqnarray}
\{\cdot,\cdot\}:\mF_{ext}^i\otimes \mF^j_{ext}\to \mF_{ext}^{i+j-1}
\end{eqnarray}
given explicitly by the expression
\begin{eqnarray}
\{a_1,a_2\}={\bf b}\mu(a_1,a_2)-\mu({\bf b}a_1,a_2)-(-1)^{n_{a_1}}\mu(a_1,{\bf b}a_2),
\end{eqnarray}
where $a_1,a_2\in \mF_{ext}^{\bf \cdot}$, also generates a homotopy Gerstenhaber algebra. We leave the question of the existence of such algebra on the 
complex $(\mF^{\bf \cdot}_{ext},\mQ)$ open, since it is rather hard to check all the relations directly. The main obstacle for its existence is hidden, of course, in the logarithmic part of the operator products which we threw away, when defined the operation $\mu$ via the OPE. We postpone the detailed study of this question to the subsequent publications. If such structure exists, it is interesting to see how it shows up in the perturbation theory of YM, e.g. in the formalism of \cite{costello}.

\section{Playing with gauge theory: other complexes and algebras.}

In the sections  2 and 3 we studied the homotopy algebras related to the canonical formulations of YM theory. Here we give some  algebraic structures related to different formulations of YM theory with and without matter fields.  \\

\noindent{\bf 4.1. First order formulation.} The Maxwell complex, which was the starting point for all the structures in this paper, contains different kinds of operators, i.e. the first and second order
ones. If one tries to break in pieces the second order operator $\ud*\ud$, one obtains the following system of maps:
\begin{eqnarray}
0\xrightarrow{ }\Omega^{0}(M)\xrightarrow{\ud}\Omega^{1}(M)\xrightarrow{*\ud}\Omega^{D-2}(M)\xrightarrow{-*}\nonumber\\
\Omega^{D-2}(M)\xrightarrow{-\ud*}\Omega^{D-1}(M)
\xrightarrow{\ud}\Omega^{D}(M)\to 0.
\end{eqnarray}
However, this is not a complex. But there is a way to turn it into a complex. We give to both $\Omega^1(M)$, $\Omega^2(M)$  and 
 $\Omega^{D-2}(M)$, $\Omega^{D-1}(M)$ the same grading and reverse the middle map, which is just the Hodge star. 
In such a way we obtain the complex 
\begin{eqnarray}
0\xrightarrow{ }\mK^0\xrightarrow{\t\mQ }\mK^1\xrightarrow{ \t\mQ}\mK^2\xrightarrow{ \t\mQ}\mK^3\xrightarrow{ }0
\end{eqnarray}
such that $\mK^0=\Omega^{0}(M)$, $\mK^{1}=\Omega^{1}(M)\oplus\Omega^{2}(M)$, $\mK^{2}=\Omega^{D-1}(M)\oplus\Omega^{D-2}(M)$, 
$\mK^{3}=\Omega^{D}(M)$ and the differential $\t \mQ$ acts as follows:
\[
\xymatrixcolsep{30pt}
\xymatrixrowsep{3pt}
\xymatrix{
0 \ar[r]&\Omega^0(M) \ar[r]^{\ud} & \Omega^1(M) \ar[ddddr]  &\Omega^{D-1}(M) \ar[r]^{\ud} & \Omega^D(M)\ar[r] & 0\\
& \quad &     & _{-{\ud}*}\quad    &&\\
 && \bigoplus & \bigoplus\quad     &&\\
 &&           & _{*{\ud}}\quad    &&\\
 && \Omega^2(M)\ar[r]_{-*} \ar[uuuur]  & \Omega^{D-2}(M)&
}
\]
This complex possess an Abelian associative graded algebra structure, such that the multiplication is defined by 
means of the following table:
\begin{center}
$\mu_{\mK}(f',f'')$=
\end{center}
\begin{tabular}{|l|c|c|c|r|}
\hline
$ f''/f'$&          $v'$ & $(\mA',{\bf F}')$ & $(\mW',{\bf G}')$ & $a'$ \\
\hline
$v''$ &                  $v'v''$    &$(v''\mA',v'' {\bf F}')$ &$(v''\mW',v''{\bf G}')$ &$v''a'$\\
\hline
$(\mA'',{\bf F}'')$     &  $(v'\mA'',v'{\bf F}'')$ & $(\mA''\wedge *{\bf F}'-$   &   $-\mA''\wedge \mW'-$&0\\
                               &                               &    $\mA'\wedge *{\bf F}'',*(\mA'\wedge\mA'')$   & ${\bf F}''\wedge {\bf G}'$&\\                    
\hline
$(\mW'',{\bf G}'')$ & $v'\mW''$ &  $-\mA'\wedge \mW''-$ &    0&0\\
&&                        ${\bf F}'\wedge {\bf G}''$&&\\
\hline
$a''$                     & $v'a''$ &   0 &    0&0\\
\hline
\end{tabular}\\

\vspace{3mm}

\noindent 
Here $f',f''$ take values in sets of variables with prime and double prime correspondingly. Other elements in the table correspond to the appropriate values of $\mu(f_1,f_2)$. In the table above $v',v''\in \mK^0$; $(\mA'',{\bf F}''), (\mA'',{\bf F}'')\in\mK^1=\Omega^1(M)\oplus \Omega^2(M)$;  $(\mW',{\bf G}'),(\mW'',{\bf G}'')\in\mK^2=\Omega^{D-1}(M)
\oplus \Omega^{D-2}(M)$;  $a',a''\in\mK^3$. 
We leave to check all the relations to the reader. 
As before, one can tensor the resulting algebra with some Lie algebra $\mg$ and obtain the graded associative algebra with a differential $\t \mQ$. The associated Maurer-Cartan equation and symmetries are:
\begin{equation}
\t Q\Phi+\mu_{\mK}(\Phi,\Phi)=0, \quad\Phi\to \Phi+\epsilon(\ud u+\mu_{\mK}(u,\Phi)-\mu_{\mK}(\Phi,u)),
\end{equation}
where $\Phi=(\mA,\mF)\in\mK^1$ and $u\in\mK^0$. One obtains that the resulting equations are 
again Yang-Mills equations with symmetries:
\begin{eqnarray}
&&\ud_{\mA}*{\bf F}=0, \quad {\bf F}=\ud\mA+\mA\wedge\mA, \nonumber\\
&&\mA\to \mA+\epsilon(\ud u+[u,\mA]),\quad  {\bf F}\to {\bf F}+\epsilon[u,{\bf F}]. 
\end{eqnarray}
It is not hard to define an invariant inner product on such an algebra, namely, it has the same form as in the usual YM theory:
\begin{eqnarray}
\langle a_1,a_2\rangle=\int_{M}a_1\wedge a_2.
\end{eqnarray}
Then, the Chern-Simons action
\begin{eqnarray}
S_{CS}=\frac{1}{2}\langle\t \mQ\Phi,\Phi\rangle +\frac{1}{3!}\langle\mu_{\mK}(\Phi,\Phi),\Phi\rangle
\end{eqnarray}
is just the first order form of YM action:
\begin{eqnarray}
S_{fo}=\frac{1}{2}Tr\int_M \Big({\bf F}\wedge*{\bf F}+*{\bf F}\wedge(\ud \mA+\mA\wedge\mA)\Big).
\end{eqnarray}
In subsection 4.2, we consider the algebraic structures related to a different first order formulation which exists 
only in four dimensions. I need to mention that so far I have not seen how this formulation can arise from CFT (see also Final remarks section for suggestions in this respect).\\

\noindent {\bf 4.2. 4D: from the first order formulation to topological YM theory.} In this subsection, the manifold $M$ will be 4-dimensional. 
In four dimensions it is possible to reduce the complex of the first order theory to the following one:
\[
\xymatrixcolsep{30pt}
\xymatrixrowsep{3pt}
\xymatrix{
0\ar[r]&\Omega^0(M) \ar[r]^\ud & \Omega^1(M) \ar[ddddr]  &\Omega^3(M) \ar[r]^\ud & \Omega^4(M)\ar[r]&0\\
 &\quad &     & _{-\ud}\quad    &&\\
 && \bigoplus & \bigoplus     &&\\
 &&           & _{\ud_{+}}\quad    &&\\
 && \Omega^2_+(M) \ar[r]_{-Id} \ar[uuuur]  & \Omega^2_+(M)&
}
\]
where $\ud_{+}=\ud+*\ud$ and $\Omega^2_+(M)$ is the space of self-dual 2-forms on the manifold $M$.  
One can define a bilinear operation on the resulting complex, slightly changing the one from the previous subsection:
\begin{center}
$\mu_{sd}(f',f'')$=
\end{center}
\begin{tabular}{|l|c|c|c|r|}
\hline
$ f''/f'$&          $v'$ & $(\mA',{\bf F}'_{+})$ & $(\mW',{\bf G}'_{+})$ & $a'$ \\
\hline
$v''$ &                  $v'v''$    &$(v''\mA',v'' {\bf F}'_{+})$ &$(v''\mW',v''{\bf G}'_{+})$ &$v''a'$\\
\hline
$(\mA'',{\bf F}''_{+})$     &  $(v'\mA'',v'{\bf F}''_{+})$ & $(\mA''\wedge {\bf F}'_{+}-$   &   $-\mA''\wedge \mW'-$&0\\
                               &                               &    $\mA'\wedge {\bf F}''_{+},P_{+}(\mA'\wedge\mA'')$   & ${\bf F}''_{+}\wedge {\bf G}'_{+}$&\\                    
\hline
$(\mW'',{\bf G}''_{+})$ & $v'\mW''$ &  $-\mA'\wedge \mW''-$ &    0&0\\
&&                        ${\bf F}'_{+}\wedge {\bf G}''_{+}$&&\\
\hline
$a''$                     & $v'a''$ &   0 &    0&0\\
\hline
\end{tabular}\\
\vspace{3mm}

\noindent Here, $P_{+}=1+*$ is the projection operator on $\Omega^2(M)$. It is not hard to see that this operation gives to the complex above the structure of Abelian associative algebra. If we tensor it with 
some Lie algebra $\mg$, following the steps of section 3, we obtain that it is related to the action
\begin{eqnarray}
S=Tr\int_M\Big({\bf F}_{+}\wedge{\bf F}_{+}+{\bf F}_{+}\wedge (\ud \mA+\mA\wedge\mA)\Big),
\end{eqnarray}
where $F\in\Omega^{+}(M)$, which is equivalent to the YM theory at least on the classical level. This algebra was also considered in \cite{costello}. 

However, using the complex above, one can obtain another very important and famous model, namely, topological Yang-Mills theory by "gluing" it with  
the Maxwell complex and the algebraic structure on it. 
One can consider the complex:
\[
\xymatrixcolsep{30pt}
\xymatrixrowsep{3pt}
\xymatrix{
0\ar[r]&\Omega^0(M) \ar[r]^{\ud} & \Omega^1(M) \ar[ddddr]\ar[r]^{\ud*\ud}  &\Omega^3(M) \ar[r]^{\ud} & \Omega^4(M)\ar[r]&0\\
 &\quad &     & _{-\ud}\quad    &&\\
 && \bigoplus & \bigoplus     &&\\
 &&           & _{\ud_{+}}\quad    &&\\
 && \Omega^2_+(M) \ar[r]_{-Id} \ar[uuuur]  & \Omega^2_+(M)&
}
\]
such that the middle square is a commutative diagram, if we reverse the $Id$ map. One can see that now it has 
both the Maxwell complex and the complex of first order theory as subcomplexes. Now one can consider the 
algebra which contains both the Abelian algebra of the first order theory and the $C_{\infty}$ algebra of YM theory as subalgebras. One can see that it is not hard to write down the explicit formulae, since operations $\mu_{sd}$ and $(\cdot,\cdot)_h$ overlap consistently.  
If we tensor it with some Lie algebra, the resulting Maurer-Cartan equation for the associated $A_{\infty}$ algebra will be just ${\bf F}=P_{+}(d\mA+\mA\wedge\mA)$.
However, this is not the end of the construction. The complex above can be extended in the following way:
\[
\xymatrixcolsep{30pt}
\xymatrixrowsep{3pt}
\xymatrix{
&& \Omega^0 \ar[r]^{\ud} & \Omega^1 \ar[ddddr]  \ar[r]^{\ud*\ud}&\Omega^3  \ar[ddddr]^{Id} \ar[r]^{\ud} & \Omega^4 \ar[ddddr]^{-Id}&&\\
&& \quad &     & _{-\ud}   &&&\\
&& \bigoplus& \bigoplus & \bigoplus     &\bigoplus&&\\
&& &           & _{\ud _+}   &&&\\
0 \ar[r]&\Omega^0 \ar[r]_{\ud} \ar[uuuur]^{-Id} &\Omega^1\ar[uuuur]^{Id}\ar[r]_{{\ud}_+} & \Omega^2_+ \ar[r]_{-Id} \ar[uuuur]  
& \Omega^{2}_+ \ar[r]_{-\ud} &\Omega^3 \ar[r]_{\ud} &\Omega^4  \ar[r]& 0
}
\]
This extension obviously corresponds to the extra symmetries (including symmetries of symmetries) of the theory. One can show that it is possible to continue the 
$C_{\infty}$ algebra nontrivially on the whole complex by means of continuation of the bilinear operation via the wedge product. 
Now, tensoring the resulting $C_{\infty}$ algebra with some Lie algebra and antisymmetrizing, one gets an $L_{\infty}$ algebra, as usual.  This algebra can  
be used to construct the BV action (as we did for YM theory in section 2), corresponding to the BV quantization of the theory with an action 
\begin{eqnarray}
S_{top}=Tr\int({\bf G}-{\bf F}_{+}(\mA))^2, \quad{\bf F}_{+}(\mA)=P_{+}(\ud\mA+\mA\wedge\mA),
\end{eqnarray}
where ${\bf G}\in\Omega^2_{+}$, which corresponds to topological YM theory (see e.g. \cite{blau}).\\

\noindent {\bf 4.3. Matter fields and Higgs mechanism.} In principle, matter fields can be added into the picture. Usually, this means that we need to extend the Maxwell complex by means 
of some contractible complex. For example, let's take the simplest nontrivial case: an Abelian gauge theory coupled to the complex scalar field. 
This means that the contractible complex should be of the form:
\begin{eqnarray}
0\to \mathfrak{R}^1 \xrightarrow{\mQ}\mathfrak{R}^2\to 0,
\end{eqnarray}
where $\mathfrak{R}^1$,   $\mathfrak{R}^2$ are the spaces of 0- and $D$-forms with values in $\mathbb{C}$ correspondingly, and the operator $\mQ$ acts as 
$\Delta+m^2$, where $\Delta$ is a Laplace operator. It is not hard to write down the relations which lead to the $L_{\infty}$ algebra.
The bilinear relation is given by:
\begin{center}$[f',f'']_h$=\end{center}

\noindent\begin{tabular}{|l|c|c|c|c|c|r|}
\hline
$ f''/f'$&$v'$                                         & $\phi'$                       &$\mA'$                                                &   $\psi'$                                          & $\mV'$      & $a'$ \\
\hline
$v''$        & $0$                                       &    $-\phi' v''$               & 0                                                         &    $-\psi' v''$                                          &  0              &0\\
\hline
$\phi''$ & $v'\phi''$                                     &    $\{\phi',\phi''\}$       &     $\{\mA',\phi''\}$                              &    $\phi''\psi'^*-\phi''^*\psi'$        &  0          &0\\
\hline
$\mA''$ &0                                                 &      $\{\mA'',\phi'\}$      & 0                                                           &    0                                               &  0               &0\\
 \hline
$\psi''$ & $v'\psi''$                                    &    $\phi'^*\psi''-\psi''^*\phi'$      &0                                          &    0                                                &  0                           & 0\\
\hline
$\mV''$ & 0                                                &      0                          &             0                                                  &     0                                                & 0&               0\\
\hline
$a'$ &    $0$                                                   &           0                       &      0                                                   &         0                                          &   0 &                 0\\
\hline
\end{tabular}\\

\vspace{3mm}

\noindent Here $v',v''\in \mF^0$; $\mA',\mA''\in\mF^1$;  $\mV'\mV''\in\mF^2$;  $a',a''\in\mF^0$; $\phi',\phi''\in \mathfrak{R}^1$; $\psi',\psi''\in \mathfrak{R}^2$. 
The bilinear forms $\{\cdot,\cdot\}$ are defined as follows:
\begin{eqnarray}
\{\mA,\phi\}&=&\ud(*\mA\phi)+\mA\wedge*\ud\phi,\nonumber\\
\{\phi',\phi''\}&=&\phi *\ud\phi'^*-\phi^**\ud\phi'+\phi'*\ud\phi^*-\phi'^*\ud\phi.
\end{eqnarray}  
The trilinear relation is given by:
\begin{eqnarray}
&&[\mA,\mB,\phi]_h=2\mA\wedge *\mB\phi, \nonumber\\
&& [\mA,\phi',\phi'']_h=-2(\phi'\phi''^*+\phi'^*\phi'')*\mA, \nonumber\\
&&[\phi',\phi'',\phi''']_h=-\lambda(\phi'\phi''\phi'''^*+
\phi'\phi''^*\phi'''+\phi'^*\phi''\phi''').
\end{eqnarray}  
Then one can check that this $L_{\infty}$ algebra is related to the action
\begin{eqnarray}
S=\int_M \Big(d_{\mA}\phi\wedge*(d_{\mA}\phi)^*+m^2\phi\phi^*-\frac{\lambda}{2}(\phi\phi^*)^2+\frac{1}{2}{\bf F}\wedge*{\bf F}\Big).
\end{eqnarray}  
The Higgs mode for this Lagrangian  can be written as a deformation of the differential $\mQ$ and the operation $[\cdot,\cdot]_h$
 given by the simplest solution $\phi_0$, such that 
$|\phi_0|^2=\frac{m^2}{\lambda}$ to the Maurer-Cartan equation
\begin{eqnarray}
\mQ\Phi+\frac{1}{2}[\Phi,\Phi]_h+\frac{1}{3!}[\Phi,\Phi,\Phi]_h=0,
\end{eqnarray} 
where $\Phi=\mA+\phi$. Namely, the deformation is given by
\begin{eqnarray}
&&\mQ_{0}\cdot=\mQ\cdot+[\phi_0,\ \cdot]_h+\frac{1}{2}[\phi_0,\phi_0,\ \cdot]_h,\nonumber\\
&&[\phi',\phi'']_{0,h}=[\phi',\phi'']_h+[\phi_0,\phi',\phi'']_h,
\end{eqnarray}
which is analogous to the detour complex which we considered in section 2. One can see that the mass generation mechanism is given by the 
deformation of the operator  $\mQ$.

In such a way, we see that this homotopic formalism we studied in this paper is quite useful and in some sense universal for the description of the 
various  aspects of field theory.

\section{Final Remarks} 
 In this paper, we studied the relations between the homotopy algebraic structures in gauge theory and in CFT. One should ask a question, whether there is such a relation in closed string case, i.e.between the homotopy structures in Gravity and possible extension of Lian-Zuckerman constructions. In \cite{zeit3}, \cite{zeit} we studied the bilinear operations, acting on the tensor product of the chiral and antichiral sectors. The operation of main interest for us is:
\begin{eqnarray}\label{bilin}
\{U^{(0)},V^{(0)}\}_{\epsilon}(z,\b z)=\mathcal{P}_{+}\int_{C_{\epsilon,z}}U^{(1)}V^{(0)}(z,\b z),
\end{eqnarray}
where $U^{(0)}$, $V^{(0)}$ are some operators on the tensor product of chiral and antichiral sectors, $U^{(1)}=\ud w[b_{-1},U^{(0)}(w,\b w)]+\ud \b w
[\t b_{-1},U^{(0)}(w,\b w)]$ 
is a 1-form, and  $C_{\epsilon,z}$ is a circle contour of radius $\epsilon$, and $\mathcal{P}_{+}$ is the projection on nonnegative powers of $\epsilon$. After 
careful analysis of this operation, one can show that it generates a Gerstenhaber algebra on the BRST cohomology, such that in the limit $\epsilon\to 0$ 
(or, replacing $\mathcal{P}_{+}$ by $\mathcal{P}_0$) this algebra reduces to the one of Lian-Zuckerman type 
extended on the product of holomorphic and antiholomorphic sectors. As we have shown in \cite{zeit3}, \cite{zeit2}, the formal Maurer-Cartan equation associated with $\epsilon\to 0$ limit of \rf{bilin}, reproduces (in quasiclassical limit) the conformal invariance conditions for the sigma model, i.e. Einstein equations, and also symmetries of these equations. We also notice that these algebraic constructions are deeply related to the very nature of the perturbation series of sigma model, regularized via (carefully defined) point-splitting. 
We will discuss in detail the operation $\rf{bilin}$ and related structures in \cite{zeitnew}. 

These CFT structures (in both open and closed string cases), which give the possibility to reproduce the classical equations of motion, should be closely related to the SFT structures, i.e. Witten's star product \cite{witten} and the Zwiebach 2-product \cite{zwiebach}. However, in the open string case the algebraic structure is very complicated, since as it was shown in \cite{taylor}-\cite{feng}, the Yang-Mills theory appears after integration out massive modes,  which gauge field is coupled to. Another problem in open string theory is on CFT side: as we have already mentioned in section 2, the Lian-Zuckerman operations should be extended (in order to satisfy the homotopy algebra relations on the full BRST complex) by means of the contributions coming from the logarithms of the OPEs. There is a way how to avoid logarithms: one can reformulate open CFT by means of some other CFT, which does not contain this complication. It can be achieved by means of consideration of the first order action, like we did for closed string case in \cite{lmz},\cite{zeit}. The first step in this direction is povided by paper \cite{fonew}, where the Yang-Mills equations were obtained from studying the Lian-Zuckerman operations in the first-order (beta-gamma) open string CFT. 

Since the Lian-Zuckerman construction appears to be very close to SFT structure, one can raise a question, whether  it is possible to reproduce "stringy" $\alpha'$-corrections directly from these algebraic structures. The first order version shows that it should be closely related to the study of the quantization of the Courant algebroid. 

In \cite{movshev} M. Movshev and A. Schwarz study possible $\alpha'$-corrections to 10-dimensional SUSY Yang-Mills theory using the vast underlying supersymmetry structure, which they incorporate into homotopy algebra language. Their approach is pure algebraic (they are relying on the pure spinor formulation) and does not involve any relation with string theory. A very natural construction would be to consider the Lian-Zuckerman structures in the context of Berkovits' formulation of superstring theory. In such a way one can compare the $\alpha'$-corrections coming directly from CFT and those, constructed algebraically.

One more question is related to the structure of noncommutative (NC) theories. 
Namely, using the introduced constructions, the $A_{\infty}$ algebra of NC Yang-Mills theory can be defined (at least in the case of 
flat space). But the intriguing question is as follows: what does Seiberg-Witten map \cite{sw} mean in these terms? The reasonable answer could be that 
it is some $A_{\infty}$-morphism, since symmetries and Maurer-Cartan equations are related in the appropriate way. 
One can also seek the relation of our constructions with the unfolded dynamics formalism (see e.g. \cite{bekaert}, \cite{vas}).
We will return to these and other 
related questions in the future.

\section*{Acknowledgements}
I would like to thank J. Block, D. Borisov, R. Donagi, I.B. Frenkel, M.M. Kapranov, M. Movshev, T. Pantev, M. Rocek, J. Stasheff, D. Sullivan and G. Zuckerman for numerous discussions on the subject. I am also  grateful to the organizers of the Simons Workshop 2008 for hospitality, support and possibility to have fruitful discussions.

\section*{{\bf A.} Proof of the homotopy algebra relations}

\noindent{\bf Proposition 3.1.} {\it Let $a_1,a_2, a_3, a_4, b, c$ $\in$ $\mF_{\mg}$. Then the following relations hold:
\begin{eqnarray}\label{rel}
&&\mQ(a_1,a_2)_h=(\mQ a_1,a_2)_h+(-1)^{n_{a_1}}(a_1,\mQ a_2)_h,\nonumber\\
&&(a_1,a_2)_h=(-1)^{n_{a_1}n_{a_2}}(a_2,a_1)_h,\nonumber\\
&&\mQ(a_1,a_2, a_3)_h+(\mQ a_1,a_2, a_3)_h+(-1)^{n_{a_1}}(a_1,\mQ a_2, a_3)_h+\nonumber\\
&&(-1)^{n_{a_1}+n_{a_2}}( a_1, a_2, \mQ a_3)_h=((a_1,a_2)_h,a_3)_h-(a_1,(a_2,a_3)_h)_h
,\nonumber\\
&&(a_1,(a_2,a_3,a_4)_h)_h+(-1)^{n_{a_1}}((a_1,a_2,a_3)_h,a_4)_h=\nonumber\\
&&((a_1,a_2)_h,a_3,a_4)_h+
(-1)^{n_{a_1}}(a_1,(a_2,a_3)_h)_h+\nonumber\\
&&(-1)^{n_{a_1}+n_{a_2}}(a_1,a_2,(a_3,a_4)_h)_h,\nonumber\\
&&((a_1,a_2, a_3)_h,b,c)_h=0.
\end{eqnarray} }

\noindent{\bf Proof.} Let's start from the first relation of Proposition 2.1.:
\begin{eqnarray}\label{der}
\mQ(a_1,a_2)_h=(\mQ a_1,a_2)_h+(-1)^{n_{a_1}}(a_1,\mQ a_2)_h.
\end{eqnarray}
We begin from the case $a_1=u\in \mF^0$. Then, for $a_2=v\in \mF^0$ we have:
\begin{eqnarray}
\mQ(\rho_u,\rho_v)_h= \ud (u,v)=(u,\ud v)_h+(\ud u, v)_h=
(\mQ u, v)_h+(u,\mQ v)_h.
\end{eqnarray}
Let  $a_2=\mA\in \mF^1$. Then 
\begin{eqnarray}\label{sum}
\mQ(u,\mA)_h=\ud *\ud (u,\mA).
\end{eqnarray}
We know that
\begin{eqnarray}
&&\m(u,\mA)=\ud *(\ud u\wedge \mA)+\ud u\wedge *\ud \mA+u\m \mA.
\end{eqnarray}
At the same time 
\begin{eqnarray}\label{1}
(\mQ u,\mA)_h=\ud u\wedge *\ud \mA+\ud*(\ud u\wedge \mA) 
\end{eqnarray}
and
\begin{eqnarray}\label{2}
(u,\mQ\mA)_h=u\m \mA.
\end{eqnarray}
Taking the sum of \rf{1} and \rf{2}, we get \rf{sum} and, therefore, the relation \rf{der} also holds in this case. The last 
nontrivial case with $a_1=u$  is that when $a_2=\mW \in \mF^2$. 
We see that 
\begin{eqnarray}
&&\mQ(u,\mW)_h=\ud (u\mW)=\ud u\wedge \mW + u\ud\mW=\nonumber\\
&&(\mQ u,\mW)_h+(u,\mQ\mW)_h.
\end{eqnarray}
Let's put $a_1=\mA \in \mF^1$. Then for $a_2=\mB \in \mF^1$, we get 
\begin{eqnarray}
\mQ(\mA,\mB)_h=\ud(\mA,\mB).
\end{eqnarray}
We find that 
\begin{eqnarray}
&&\ud(\mA,\mB)=(\ud \mA\wedge*\ud \mB)-(\mA\wedge\ud*\ud\mB)+\nonumber\\
&&(\mB\wedge\ud*\ud\mA)-
(*\ud\mA\wedge\ud\mB)=(\ud*\ud\mA,\mB)_h-(\mA,\ud*\ud\mB)_h.
\end{eqnarray}
This leads to the relation:
\begin{eqnarray}
\mQ(\mA,\mB)=(\mQ\mA,\mB)_h-(\mA,\mQ\mB)_h,
\end{eqnarray}
where $\{\mA,\mB\}\in \mF^1$. Therefore, \rf{der} holds in this case. 

It is easy to see that this relation for the other values of 
$a_1$ and $a_2$ reduces to trivial one. Thus, we proved \rf{der}, i.e. the first relation from Proposition 2.1.
The graded commutativity condition $(a_1,a_2)_h=(-1)^{n_{a_1}n_{a_2}}(a_2,a_1)_h$ follows directly from the definition.
The next relation to prove is 
the deformed associativity condition
\begin{eqnarray}\label{assoc}
&&\mQ(a_1,a_2, a_3)_h+(\mQ a_1,a_2, a_3)_h+(-1)^{n_{a_1}}(a_1,\mQ a_2, a_3)_h+\nonumber\\
&&(-1)^{n_{a_1}+n_{a_2}}( a_1, a_2, \mQ a_3)_h=((a_1,a_2)_h,a_3)_h-(a_1,(a_2,a_3)_h)_h.
\end{eqnarray}
It is easy to see that due to the graded commutativity property of the bilinear operation it is enough to prove it in two cases: i)$a_1\in\mF^0$, $a_2,a_3\in\mF^1$
and ii) $a_1,a_2,a_3\in\mF^1$. Let's show i) at first.
\begin{eqnarray}
\big(u,(\mA,\mB)_h\big)_h=u\mA\wedge*\ud \mB-u\mB\wedge*\ud \mA+u\ud*(\mA\wedge\mB),\nonumber\\
\big((u,\mA)_h,\mB\big)_h=u\mA\wedge*\ud \mB-u\mB\wedge*\ud (u\mA)+u\ud*(u\mA\wedge\mB).
\end{eqnarray}
Therefore 
\begin{eqnarray}
&&\big((u,\mA)_h,\mB\big)_h-\big(u,(\mA,\mB)_h\big)_h=\nonumber\\
&&\ud u\wedge*(\mA\wedge\mB)-\mB\wedge *(\ud u\wedge\mA)=(\ud u,\mA,\mB)_h.
\end{eqnarray}
Thus i) is proven. Now we  prove ii).
 \begin{eqnarray}
&& \big((\mA,\mB)_h,\mC\big)_h=\mC\wedge\mA\wedge*\ud \mB-\mC\wedge\mB\wedge*\ud \mA+\mC\wedge\ud*(\mA\wedge\mB)\nonumber\\
 &&=\mC\wedge\mA\wedge*\ud \mB-\ud \mA\wedge*(\mC\wedge\mB)+\mC\wedge\ud*(\mA\wedge\mB), \nonumber\\
 &&\big(\mA,(\mB,\mC)_h\big)_h=\mA\wedge\mB\wedge*\ud \mC-\mA\wedge\mC\wedge*\ud \mB+\mA\wedge\ud*(\mB\wedge\mC)\nonumber\\
 &&=\ud \mC\wedge*(\mA\wedge\mB)-\mA\wedge\mC\wedge*\ud \mB+\mA\wedge\ud*(\mB\wedge\mC).
 \end{eqnarray}
 Hence, 
 \begin{eqnarray}
\big((\mA,\mB\big)_h,\mC)_h-\big(\mA,(\mB,\mC)_h\big)_h=\ud(\mA,\mB,\mC)_h.
 \end{eqnarray}
Thus \rf{assoc} is proven.
The next relation to prove is the generalized associativity condition
\begin{eqnarray}\label{genassoc}
&&\big(a_1,(a_2,a_3,a_4)_h\big)_h+(-1)^{n_{a_1}}\big((a_1,a_2,a_3)_h,a_4\big)_h=\nonumber\\
&&\big((a_1,a_2)_h,a_3,a_4\big)_h
+(-1)^{n_{a_1}}\big(a_1,(a_2,a_3)_h\big)_h+\nonumber\\
&&(-1)^{n_{a_1}+n_{a_2}}\big(a_1,a_2,(a_3,a_4)_h\big)_h.
 \end{eqnarray}
 Easy to see that it is enough to examine two cases: i) all $a_i\in\mF^1$, ii) one of $a_i$ belongs to $\mF^0$, all other belong to $\mF^1$. i) follows from the relation
 \begin{eqnarray}
\big((\mD,\mA,\mB)_h,\mC\big)_h-\big(\mD,(\mA,\mB,\mC)_h\big)_h=0,
 \end{eqnarray}
and ii) follows from
\begin{eqnarray} 
 \big((u,\mA)_h,\mB,\mC\big)_h=\big(u,(\mA,\mB,\mC)_h\big)_h.
\end{eqnarray}
The last relation we need to prove is for trilinear operation only:
\begin{eqnarray}
\big((a_1,a_2, a_3)_h,b,c\big)_h=0.
\end{eqnarray}
It is trivial to check due to the fact that the trilinear operation is nonzero only if all arguments belong to 
$\mF^1$. 
Thus Proposition 3.1. is proven. $\blacksquare$\\

\noindent{\bf Proposition 3.3.} 
{\it The operations $\mu, n,m$ satisfy the following relations:
\begin{eqnarray}
&&\mQ\mu(a_1,a_2)=\mu(\mQ a_1,a_2)+(-1)^{n_{a_1}}\mu(a_1,\mQ a_2),\nonumber\\
&&\mu(a_1,a_2)-(-1)^{n_{a_1}n_{a_2}}\mu(a_2,a_1)=\nonumber\\
&&\mQ m(a_1,a_2)+m(\mQ a_1,a_2)+(-1)^{n_{a_1}}m(a_1,\mQ a_2)\nonumber\\
&&\mu\big(\mu(a_1,a_2),a_3\big)-\mu\big(a_1,\mu(a_2,a_3)\big)=\nonumber\\
&&\mQ n(a_1,a_2,a_3)+n(\mQ a_1,a_2,a_3)+(-1)^{n_{a_1}}n(a_1,\mQ a_2,a_3)+\nonumber\\
&&(-1)^{n_{a_1}n_{a_2}}n(a_1,a_2,\mQ a_3),
\end{eqnarray}
where $a_i\in\mF^{\bf \cdot}_{ext}$.}\\

\noindent{\bf Proof.} Let's prove the first relation, i.e. we prove that $\mQ$ is a derivation for $\mu$. We check the relations 
which do not trivially follow from the  relations of Proposition 3.1:
\begin{eqnarray}
&&\mu(\mQ u,\mv)+\mu(u,\mQ\mv)=*\ud u\mv+u*\mv-*(\ud u \mv)=*u\mv=\mQ\mu(u,\mv),\nonumber\\
&&\mu(\mQ\mA,\mv)-\mu(\mA,\mQ\mv)=\ud*(\mA\mv)=\mQ\mu(\mA,\mv),\nonumber\\
&&\mu(\mQ u,\ma)+\mu(u,\mQ\ma)=-\ud*(\mA\wedge*\ma)=\mQ\mu(u,\ma).
\end{eqnarray} 
In the formulae above $u\in \mF^{0}$, $\mA\in \mF^1$, $\mv\in \mG^1$, $\ma\in\mG^2$. 
The second relation (the commutativity of $\mu$ up to homotopy) follows directly from the explicit form of $\mu$ and $n$. The last relation we need to prove 
is:
\begin{eqnarray}\label{assmu}
&&\mu\big(\mu(a_1,a_2),a_3\big)-\mu\big(a_1,\mu(a_2,a_3)\big)=\nonumber\\
&&\mQ n(a_1,a_2,a_3)+n(\mQ a_1,a_2,a_3)+(-1)^{n_{a_1}}n(a_1,\mQ a_2,a_3)+\nonumber\\
&&(-1)^{n_{a_1}n_{a_2}}n(a_1,a_2,\mQ a_3).
\end{eqnarray} 
Again, we check only the nontrivial relations. First, we prove \rf{assmu} when all arguments belong to $\mF^1$:
\begin{eqnarray}
&&\mu(\mu(\mA,\mB),\mC)-\mu(\mA,\mu(\mB,\mC))=\nonumber\\
&&((\mA,\mB)_h,\mC)_h-(\mA,(\mB,\mC)_h)_h+\mu(\mA\wedge*\mB,\mC)-\mu(\mA,\mB\wedge *\mC)=\nonumber\\
&&-\ud*(\mC\wedge *(\mA\wedge *\mB))+\ud*(\mA\wedge *(\mB\wedge *\mC)) +\mQ(\mA,\mB,\mC)_h=\nonumber\\
&&\ud(*\mA\wedge*(\mB\wedge\mC)-*\mC\wedge*(\mA\wedge*\mB)+\nonumber\\
&&\mA\wedge*(\mB\wedge\mC)-\mC\wedge*(\mA\wedge\mB))=\mQ n(\mA,\mB,\mC).
\end{eqnarray} 
Another nontrivial case is when one of the arguments belongs to $\mG^1$:
\begin{eqnarray}
&&\mu(\mu(\mA,\mB),\mv)-\mu(\mA,\mu(\mB,\mv))=\mu(\mA\wedge *\mB,\mv)-\mu(\mA,*\mB\mv)=0,\nonumber\\
&&\mu(\mu(\mA,\mv),\mB)-\mu(\mA,\mu(\mv,\mB))=\mu(*\mA\mv,\mB)+\mu(\mA,*\mB\mv)=\nonumber\\
&&2*(\mA\wedge *\mB\mv)=-m(\mA,\mQ\mv,\mB),\\
&&\mu(\mu(\mv,\mA),\mB)-\mu(\mv,\mu(\mA,\mB))=-2*(\mA\wedge *\mB)\mv=m(\mQ\mv,\mA,\mB).\nonumber
\end{eqnarray} 
And the last nontrivial situation is when one of the arguments belongs to $\mF^0$, another to $\mG^2$ and  the last one to $\mF^1$:
\begin{eqnarray}
&&\mu(\mu(u,\ma),\mA)-\mu(u,\mu(\ma,\mA))=\mu(\ma u-*(\ud u\wedge *\ma),\mA)+\nonumber\\
&&\mu(u,\ud*(\mA\wedge *\ma))= -\ud*(\mA\wedge *(au))-\mA\wedge*(\ud u\wedge*\ma)+\nonumber\\
&&u\ud*(\mA\wedge *\ma)=-2\mA\wedge*(\ud u\wedge*\ma)=(\mQ u, \ma,\mA),\nonumber\\
&&\mu(\mu(\ma,u),\mA)-\mu(\ma,\mu(u,\mA))=-\ud*(\mA\wedge *(au))-\mA\wedge*(\ud u\wedge*\ma)+\nonumber\\
&&\ud*(\mA\wedge *u\ma)-\ma\wedge *(\ud u\wedge *\mA)=-2\mA\wedge*(\ud u\wedge*\ma)=(\ma,\mQ u, \mA),\nonumber\\
&&\mu(\mu(\mA,\ma),u)-\mu(\mA,\mu(\ma,u))=-u\ud*(\mA\wedge *\ma)+\ud*(\mA\wedge *u\ma)+\nonumber\\
&&\mA\wedge *(\ud u\wedge *\ma)=-m(\mA,\ma,\ud u),\nonumber\\
&&\mu(\mu(u,\ma),\mA)-\mu(u,\mu(\ma,\mA))=\mu(u\ma-*(\ud u\wedge *\ma),\mA)+\nonumber\\
&&u\ud *(\mA\wedge *\ma)= 
-2\mA\wedge*(\ud u\wedge*\ma)=(\mQ u,\ma, \mA),\nonumber\\
&&\mu(\mu(u,\mA),\ma)-\mu(u,\mu(\mA,\ma))=\mu(u\mA+*(\ud u\wedge *\mA), \ma)+\nonumber\\
&&u\ud *(\mA\wedge *\ma)=0,\nonumber\\
&&\mu(\mu(\mA,u),\ma)-\mu(\mA,\mu(u,\ma))=\mu(\mA u-*(\ud u\wedge*\mA),\ma)-\nonumber\\
&&\mu(\mA, u\ma-*(\ud u\wedge *\ma))=-\ud*(\mA\wedge u\ma)-*(\ud u\wedge*\mA)\ma+\nonumber\\
&&\ud*(\mA\wedge u\ma)+\mA\wedge*(\ud u\wedge *\ma)=0.
\end{eqnarray} 
This finishes the Proof. $\blacksquare$

\section*{{\bf B.} Lian-Zuckerman operations: explicit computations}

In this Appendix, we will calculate explicitly the leading $\alpha'$-corrections to the operations $\mu$, $n$, $m$ in the open string case. We we will show how these OPE formulas lead to the table from subsection 3.5.   

We remind that in subsection 3.2. we associated 
$u\in \mF^0$ with $\rho_u=u(X)$, $\mau\in\mG^1$ with $\phi_{\mau}=2\alpha'\p c $,  
$\mA\in\mF^1$ with $2\phi_{\mA}=2cA_{\mu}(X)\p X^{\mu}-2\alpha'\p c \p_{\mu}A^{\mu}$, 
$\ma \in\mG^2$ with $2\psi_{\ma}=2\alpha' c\p^2 c\ma(X)+4 \alpha'c\p c \p_{\mu}\ma(X)\p X^{\mu}$, 
$\mW\in\mF^2$ with $4\psi_{\mW}=4\alpha' c\p c \mW_{\mu}\p X^{\mu}$, 
 $a\in\mF^3$ with $-4\chi_a=-4\alpha'^2 c\p c \p^2 c$. 

Now, we will show how the leading $\alpha'$ orders for the operation $\mu(\cdot, \cdot)$ on vertex operators induce bilinear operation on fields. Let's start with the first nontrivial case, when the arguments are $\rho_u$ and $\phi_{\mA}$:
\begin{eqnarray}
&&\mu(\rho_u,\phi_{\mA})=c uA_{\mu}\p X^{\mu}-\alpha'\p c u\p_{\mu}A^{\mu}=\nonumber\\
&&\phi_{u\mA}+\alpha'\p c \p_{\mu}A^{\mu}=\phi_{u\mA}+\phi_{*(\ud u\wedge *\mA)},\nonumber\\
&&\mu(\phi_{\mA}, \rho_u)=cA_{\mu}u\p X^{\mu}-\alpha' \p c \p_{\mu}A^{\mu} u-2\alpha' \p c A_{\mu}\p^{\mu}u=\nonumber\\
&&\phi_{u\mA}-\phi_{*(\ud u\wedge *\mA)}.
\end{eqnarray} 
Another nontrivial example occurs in the case when both arguments are 1-forms: 
\begin{eqnarray}
&&\mu(\phi_{\mA},\phi_{\mB})=\alpha'(-\frac{1}{2}\p^2 c c A^{\mu}B_{\mu}-2\p c c A_{\mu}\p^{\mu}B_{\nu}\p X^{\nu}+\nonumber\\
&&2\p c c\p^{\rho}A_{\mu}B_{\rho}\p X^{\mu}-2\p c c\p_{\rho}A_{\mu}B^{\mu}\p X^{\rho}-\p c c \p_{\mu} A^{\mu}B_{\rho}\p X^{\rho}- \nonumber\\
&&c\p c \p_{\mu}B^{\mu}A_{\rho}\p X^{\rho})=\psi_{(\mA,\mB)}+\frac{1}{2} \psi_{\mA\wedge *\mB}, 
\end{eqnarray} 
and  $(D-1)$-, $D$-forms:
\begin{eqnarray}
&&\mu(\rho_u,\psi_{\ma})=\alpha' c\p^2 c u\ma+\alpha' c\p c u\p_{\mu}\ma\p X^{\mu}=\psi_{u\ma}-2\psi_{*(du\wedge *\ma)}\nonumber\\
&&\mu(\phi_{\mA},\psi_{\mW})=-\frac{1}{2}\alpha'^2c\p c \p^2 cA_{\mu}W^{\mu}=-\frac{1}{2}\chi_{\mA\wedge *\mW},\nonumber\\
&&\mu(\psi_{\mW}, \phi_{\mA})=-\alpha'^2(\frac{1}{2} \p^2 c\p c c W_{\mu}A^{\mu}+\p c \p^2 c c W_{\mu}A^{\mu})=\nonumber\\
&&-\frac{1}{2}\chi_{\mA\wedge*\mW}.
\end{eqnarray}
In such a way one can fill all the table of section 3.5. We leave all other cases as an exercise. 
Similarly, one can show that operation $m$ gives nonzero value only when its arguments correspond to 1-forms. And in this situation we have
\begin{eqnarray}
m(\phi_{\mA},\phi_{\mB})=-b_{-1}\alpha' \p c c A_{\mu}B^{\mu}=\alpha'\p c A_{\mu}B^{\mu}=\phi_{*(\mA\wedge *\mB)},
\end{eqnarray}
which is in agreement with \rf{mexpl}. Finally, for the trilinear operation $n(\cdot, \cdot, \cdot)$ we get:
\begin{eqnarray}
&&n(\phi_{\mA},\phi_{\mB},\phi_{\mC})=c\p c (A_{\nu}B_{\mu}C^{\mu}-B_{\nu}A_{\mu}C^{\mu})\p X^{\nu}=\nonumber\\
&&\frac{1}{2}
\psi_{*\mA\wedge*(\mB\wedge *\mC)-*\mC\wedge*(\mA\wedge*\mB)+\mA\wedge*(\mB\wedge\mC)-\mC\wedge*(\mA\wedge\mB)}\nonumber\\
&&n(\psi_{\ma},\phi_{\mA},\phi_{\mB})=-\p^2 c\p c c A_{\mu}B^{\mu}=2\chi_{\ma\wedge *(\mA\wedge *\mB)},
\end{eqnarray}
and it vanishes in the cases when arguments are outside $\mathcal{F}^1, \mathcal{G}^2$. This is in agreement with \rf{nexpl}.

\end{document}